\documentclass[letterpaper,11pt]{article}
\pdfoutput=1

\usepackage{url}
\usepackage{jheppub}
\usepackage{multirow}
\usepackage{subfig}
\usepackage{xspace}
\usepackage{xcolor}
\usepackage[countmax]{subfloat}
\usepackage{amsmath}
\usepackage{mathtools}
\usepackage{graphicx}
\usepackage{slashed}
\usepackage{tikz-feynman,contour}
\usepackage{tikz,pgf}
\usepackage{braket}
\usepackage{amsfonts}
\usepackage{amsthm,scrextend}
\usepackage{comment}
\usepackage{placeins}
\allowdisplaybreaks

\newcommand{\as}{\alpha_S}
\newcommand{\cf}{C_{\text{F}}}
\newcommand{\ca}{C_{\text{A}}}
\newcommand{\nf}{n_{\text{F}}}

\newcommand{\msb}{\overline{\text{MS}}}

\newcommand{\kt}{k_t}

\newcommand{\mur}{\mu_R}

\newcommand{\ord}[1]{{\cal O}\!\left(#1\right)}
\newcommand{\EMD}{\text{EMD}}
\usepackage{color}
\definecolor{darkblue}{rgb}{0,0,0.5}
\definecolor{darkgreen}{rgb}{0,0.5,0}
\definecolor{darkorange}{rgb}{0.8,0.3,0}

\newcommand{\eventtwo}{\texttt{EVENT2}}

\title{Event isotropy in perturbative QCD}

\author[1,2,3]{Daniele Atzori,}
\author[1,2]{Matteo Cacciari,}
\author[3]{Simone Marzani,}
\author[4]{Gregory Soyez}
\affiliation[1]{Sorbonne Universit\'e, CNRS, Laboratoire de Physique Th\'eorique et Hautes \'Energies,
LPTHE, F-75005 Paris, France}
\affiliation[2]{Universit\'e Paris Cit\'e, LPTHE, F-75006 Paris, France}
\affiliation[3]{Dipartimento di Fisica, Universit\`a di Genova and INFN, Sezione di Genova, Via Dodecaneso 33, 16146, Italy}
\affiliation[4]{Institut de Physique Th\'{e}orique, Paris Saclay University, CNRS, CEA,
Orme des Merisiers, B\^{a}t 774, F-91191. Gif-sur-Yvette, France}

\emailAdd{atzori@lpthe.jussieu.fr}
\emailAdd{cacciari@lpthe.jussieu.fr}
\emailAdd{simone.marzani@ge.infn.it}
\emailAdd{gregory.soyez@ipht.fr}

\abstract{
It has recently been proposed that collider events can be equipped with a metric, the Energy Mover's Distance (EMD), which allows one to rephrase multiple aspects of collider physics in a geometric language. Further, the \EMD~can be exploited to define new observables. In this context, event isotropy quantifies the resemblance of an event to a uniform energy distribution. 
We present the first theoretical study of event isotropy within the framework of perturbative QCD. 
In particular, we first obtain the isotropy distribution in $e^+e^-$ collisions 
semi-analytically at ${\cal O}(\as)$ (leading-order) and numerically at ${\cal O}(\as^2)$ (next-to-leading order, NLO). 
Furthermore, we obtain all-order theoretical predictions by resumming the contributions of soft and collinear emissions, at next-to-leading logarithmic (NLL) accuracy. By matching resummed and fixed-order predictions we reach NLL+NLO accuracy.
}

\begin{document}
\maketitle

\section{Introduction}\label{sec:intro}

The Large Hadron Collider (LHC) at CERN probes particle interactions at the smallest experimentally accessible distance scales. Proton--proton collisions at a centre-of-mass energy of 13.6~TeV produce complex final states, typically containing hundreds or even thousands of particles. A central task of collider physics is to describe these final states with sufficient accuracy to disentangle rare processes from backgrounds that are often orders of magnitude larger. Meeting this challenge requires a detailed understanding of QCD dynamics, together with analysis techniques capable of exploiting the full structure of the data.

Jets---collimated sprays of hadrons---are among the most common and important objects in LHC events. The theoretical description of jets has progressed steadily over the past decades~\cite{Marzani:2019hun}, driven by advances in perturbative QCD at fixed order and by the resummation of logarithmically enhanced contributions associated with soft and collinear radiation.

At present, however, the theoretical precision for many jet observables is becoming a limiting factor. Experimental uncertainties continue to decrease, while perturbative uncertainties often remain at the level of several percent or more. To remain competitive, theoretical predictions must reach next-to-next-to-leading order (NNLO) in the strong coupling and be matched to resummation at next-to-next-to-leading logarithmic (NNLL) accuracy. This is particularly relevant for jet substructure observables, which are sensitive to multiple scales and play an increasingly important role in LHC analyses. Extending NNLL-accurate predictions to jet substructure remains a challenge because soft gluon radiation in events with multiple resolution scales gives rise to the intricate structure of non-global~\cite{Dasgupta:2001sh,Banfi:2021owj,Banfi:2021xzn,Becher:2016mmh,Balsiger:2018ezi,Becher:2023vrh} and clustering logarithms~\cite{Banfi:2005gj,Delenda:2006nf,Becher:2023znt,Delenda:2012mm,Khelifa-Kerfa:2024dut,Khelifa-Kerfa:2025jzl}. 
Improved perturbative accuracy also places greater demands on the modelling of non-perturbative effects, including hadronisation, multiple-parton interactions, and pileup. 

Alongside these developments, machine-learning techniques have become an integral part of collider physics. In jet physics, in particular, machine-learning-based classifiers have achieved significant performance gains in tasks such as flavour tagging and signal--background discrimination. While these methods are highly effective, it is important to understand which physical features they exploit, how robust they are to theoretical and experimental uncertainties, and how their behaviour can be related to well-defined quantities in quantum field theory.

This has motivated increased interest in theory-driven approaches to classification. Central concepts include the likelihood ratio, which defines the optimal classifier for a given problem, and observables that quantify distances or similarities between collision events. A natural way to characterise events is through their energy flow, describing how energy is distributed across directions in the detector.~\footnote{This observation is the starting point of an alternative, and equally successful, approach to jet substructure that exploits energy correlators rather than binned observables, see~\cite{Moult:2025nhu} and references therein.}
From this viewpoint, two events may be regarded as similar if their energy flows are similar. This idea can be formalised by equipping the space of events with a metric.

In refs.~\cite{Komiske:2019fks,Komiske:2020qhg}, such a metric was introduced using ideas from Optimal Transport theory. The resulting Energy Mover's Distance (\EMD) measures the minimal amount of ``work'' required to rearrange the energy flow of one event into that of another. The \EMD~is a proper metric and, under certain assumptions, it coincides with the Wasserstein distance~\cite{kantorovich1942translocation,wasserstein1969markov} widely used in optimal transport. Equipping collider events with a metric structure allows for a geometric interpretation of familiar observables. One can define manifolds of idealised events and interpret event shapes as distances to these manifolds.
From this perspective, the thrust observable~\cite{Farhi:1977sg}, which is one of the most-studied observables in electron-positron ($e^+e^-$) collisions, can be understood as the \EMD~between a given event and the manifold of perfectly two-jet-like events consisting of two back-to-back particles. Similarly, event isotropy~\cite{Cesarotti:2020hwb} is defined as the \EMD~between an event and an idealised configuration with uniform radiation.

Event isotropy is of phenomenological interest for searches involving approximately iso\-tropic final states~\cite{Cesarotti:2020uod,Cesarotti:2020ngq} and for mitigating the effects of pileup, which will pose a major challenge at the High-Luminosity LHC~\cite{Cesarotti:2020xtf}.
Event isotropy has recently been measured by the ATLAS collaboration~\cite{ATLAS:2023mny} and compared to predictions from standard Monte Carlo event generators. To date, however, no first-principle theoretical predictions for this observable exist. This stands in contrast to classic event shapes observables, which have been studied extensively in perturbative QCD, e.g.~\cite{Catani:1992ua,Banfi:2004yd,Banfi:2014sua,Hoang:2025uaa,Abbate:2010xh,Hoang:2014wka,Aglietti:2025jdj}.

Providing a framework for the calculation of \EMD-based observables from first principles is therefore an important open problem. 
For instance, it was recently suggested that spectral versions of the \EMD~\cite{Larkoski:2023qnv} are more amenable to study in QCD because one can obtain closed-form expressions~\cite{Gambhir:2024ndc}, thus enabling their high-precision determination through resummation~\cite{Larkoski:2025pai}.
In this work, we follow a different path and develop a theoretical framework for the perturbative calculation of observables defined through the Energy Mover's Distance, focusing on the case of event isotropy in $e^+e^-$ collisions. 
Our choice is motivated not only by the fact that $e^+e^-$ collisions provide a simpler theoretical environment in which to establish the perturbative structure of event isotropy, but also by the renewed interest in exploiting archived LEP data for precision QCD studies~\cite{Electron-PositronAlliance:2019cpi,Electron-PositronAlliance:2021kig,Electron-PositronAlliance:2023klx,Electron-PositronAlliance:2025hze,Electron-PositronAlliance:2025fhk}. 

The paper is organised as follows. In section~\ref{sec:definitions} we introduce event isotropy, while in section~\ref{sec:observable} we develop a strategy to solve the optimal transport problem that defines the observable in a semi-analytic way. We then compute perturbative predictions for the event-isotropy distribution in section~\ref{sec:QCD}, exploiting fixed-order techniques and resummation, before drawing our conclusions in section~\ref{sec:conclusions}. Technical details are collected in the appendices.

\section{Definition of event isotropy}\label{sec:definitions}
Taking inspiration from the theory of optimal transport, the authors of refs.~\cite{Komiske:2019fks,Komiske:2020qhg} adapted the notion of the Earth Mover's Distance to the description of states resulting from high-energy particle collisions.
In particular, they introduced the Energy Mover's Distance (\EMD), which measures the ``distance'' between two collider events as the minimum amount of ``work'' required to rearrange one collider event into the other.

Consider two events with the same total energy, event $\mathcal{E}$ with $M$ particles and event $\mathcal{E'}$ with $M'$ particles. Let $\omega_i$ and $\omega_j'$ denote kinematical weights characterising the $i$-th particle of $\mathcal{E}$ and the $j$-th particle of $\mathcal{E'}$, respectively. These weights are normalised such that $\sum_i\omega_i=\sum_j\omega'_j=1$. The \EMD~is defined as 
\begin{equation}\label{norm_EMD}
	\EMD_{\beta}\left(\mathcal{E},\mathcal{E'}\right)=\underset{\left\{f_{ij}\geq0\right\}}{\min}\,\sum_{i=1}^{M}\sum_{j=1}^{M'}f_{ij}d_{ij}^{\beta},
	\end{equation}
	\\
	with the constraints~\footnote{More generally, one can define a distance between events with different energies. In such a case, the constraints in eq. (\ref{eq:constraints}) generalise to inequalities, see~\cite{Komiske:2020qhg}.}
	\begin{equation}
    \label{eq:constraints}
f_{ij}\geq0, \qquad \sum_{i=1}^{M}f_{ij}=\omega_j', \qquad \sum_{j=1}^{M'}f_{ij}=\omega_i, \qquad \sum_{i=1}^{M}\sum_{j=1}^{M'}f_{ij}=1.
	\end{equation}
The so-called ground metric $d_{ij}$ is the distance between particles $i$ and $j$, where $\beta$ is a positive parameter. The quantities $f_{ij}$ represent the transportation plan and describe the fraction of energy moved from particle $j$ to particle $i$. 
The \EMD~is positive, symmetric, and it is easy to see that $\EMD_{\beta} (\mathcal E, \mathcal E)=0$.
\footnote{Furthermore,  it can be shown that $\EMD_{\beta}^{1/\beta}$ satisfies the triangle inequality: it is therefore a metric. 
In such a case, $\EMD_{\beta}^{1/\beta}$ coincides with the $p$-Wasserstein metric~\cite{kantorovich1942translocation,wasserstein1969markov}, with $p=\beta$.}
The minimisation in eq.~(\ref{norm_EMD}) defines an optimal transport (OT) problem. Accordingly, the \EMD~corresponds to the minimal work needed to rearrange one event into the other by redistributing the whole energy distribution of one event to the other.

The event isotropy $\mathcal{I}_n$ of an event $\mathcal{E}_n$ with $n$ particles can be defined as the distance between this event and an idealised isotropic event $\mathcal{U}$ with the same total energy~\cite{Cesarotti:2020hwb}:
\begin{equation} \label{isotropy}
\mathcal{I}_n^{\text{geo}}(\mathcal{E}_n)=\text{\EMD}_{\beta}\left(\mathcal{E}_n,\mathcal{U^{\text{geo}}}\right),
\end{equation}
where $\mathcal{U^{\text{geo}}}$ consists of a continuous distribution of particles with a uniform energy density over a particular geometry. In such a case, the optimal transport problem is referred to as a semi-discrete one. 

Spherical geometry is appropriate to study final-state events of $e^+e^-$ collisions. Ring-like or cylindrical geometries, instead, are useful for dealing with $pp$ collisions, or, in general, processes where transverse components of the momenta of the particles are singled out.
In this paper, we concentrate on $e^+e^-$ collisions, and we therefore consider a spherical geometry, omitting the specific label  (geo=sphere) in eq.~(\ref{isotropy}).
In this context, the weights of the particles and the Euclidean distances between them are
\begin{equation} \label{weights_distances_sph}
		\omega_i=\frac{E_i}{Q}, \qquad d_{ij}=\sqrt{\left|\frac{\vec{p}_i}{|\vec{p_i}|} - \frac{\vec{p}_j}{|\vec{p_j}|}\right|^2} =\sqrt{2\left(1-\cos\theta_{ij}\right)},
\end{equation}
where $Q$ is the total c.o.m. energy, $E_i$ and $\vec{p_i}$ are the energy and the spatial momentum of particle $i$, respectively, and $\theta_{ij}$ is the angle between the directions of particles $i$ and $j$.
Note that, with our definition in eq.~(\ref{norm_EMD}), event isotropy  is  normalized to lie in $[0,1]$: $\mathcal{I}=0$ (most isotropic) for a uniformly distributed event and $\mathcal{I}=1$ (least isotropic) for a back-to-back two-particle event.\footnote{This definition of isotropy ${\cal I}$ is counterintuitive, as its value is smaller for the more isotropic events. In the following we will work with $\iota \equiv 1 - {\cal I}$, and sometimes this quantity will be referred to as the isotropy.} 
Finally, we work with the exponent $\beta=2$.
    	
\section{Semi-analytic determination of event isotropy}\label{sec:observable}
In this section, we discuss a general strategy for solving the OT problem that defines event isotropy. We then apply this strategy to the case of final states with two or three particles, obtaining closed-form semi-analytic expressions for event isotropy in these two cases.
	
Let us start by considering an $n$-particle final state event $\mathcal{E}_n\left(\left\{p_i\right\}\right)$, where $p_i$ are the particles' four-momenta, and a uniform spherical event $\mathcal{U}$, described by a uniform energy density distribution over the $S^2$ sphere
	\begin{equation}
		\omega'\left(\Omega'\right)=\frac{1}{4\pi},
	\end{equation}
where $\Omega'\equiv\left(\theta',\phi'\right)$ identifies a point on the sphere through its polar and azimuthal angles. 

The isotropy of $\mathcal{E}_n$ is the solution of a semi-discrete optimal transport (SDOT) problem
	\begin{equation}
		\label{Insph_def}
		\mathcal{I}_n\left(\left\{p_i\right\}\right)=\text{EMD}_2\left(\mathcal{ E}_n\left(\left\{p_i\right\}\right),\mathcal{U}\right)=\underset{\left\{f_i\right\}}{\min}\,\sum_{i=1}^{n}\int_{S^2}d\Omega'\,f_i(\Omega')\,d_i^2(\Omega'),
	\end{equation}
with the constraints
	\begin{equation}
		\label{constraints_semi_D}
\int_{S^2}d\Omega'\,f_i(\Omega')=\omega_i,\hspace{10ex}\sum_{i=1}^{n}f_i(\Omega')=\omega'\left(\Omega'\right)=\frac{1}{4\pi}.
	\end{equation}
Note that the transportation plan is now a discrete set of functions defined over the unit sphere: $f_{ij}\rightarrow{}f_i(\Omega')$, and  the same applies for the distances: $d_{ij}\rightarrow{}d_i(\Omega')$. $d_i^2(\Omega')$ is the squared distance between the $i$-th particle of event $\mathcal{E}_n$ and the ``particle" of $\mathcal{U}$ at point $\left(\theta',\phi'\right)$, and $f_i(\Omega')$ represents the fraction of energy density moved from there to the $i$-th particle of $\mathcal{E}_n$.

An important result from the theory of OT, see e.g.~\cite{santambrogio2015optimal,peyre2020computationaloptimaltransport}, is the fact that the solution of a semi-discrete OT problem with a continuous uniform density distribution and a Euclidean ground distance is a piecewise transportation plan $f_i$ on the so-called Laguerre cells $C_i$, partitions of the sphere that generalise the concept of Voronoi cells, 
	\begin{equation}
		\label{transportation_plan_Laguerre}
		f_i\left(\Omega'\right)=\left\{
		\begin{array}{ll}
			1/(4\pi) & \text{if}\hspace{2ex}\Omega'\in{}C_i
			\\
			0 & \text{otherwise}
		\end{array}
		\right..
	\end{equation}
To be more precise, given a set of $N$ sites (points) $x_1,...,x_N$ in a metric space $\left(X,d\right)$, the Voronoi cell $V_i$ associated with the site $x_i$ is the region of the space consisting of all points closer to $x_i$ than to any other sites:
	\begin{equation}
		V_i=\left\{\left.y\in{}X\hspace{1ex}\right|\hspace{1ex}d\left(x_i,y\right)-d\left(x_j,y\right)\leq0\right\}.
	\end{equation}
The Laguerre cells $C_i$ are weighted Voronoi cells:
		\begin{equation}
		\label{Laguerre_def}
		C_i=\left\{\left.y\in{}X\hspace{1ex}\right|\hspace{1ex}d\left(x_i,y\right)-d\left(x_j,y\right)\leq{}c_{ij}\right\},
	\end{equation}
where $c_{ij}$ are constants depending on $i$ and $j$.
Unlike Voronoi ones, Laguerre cells are not uniquely determined by the particle positions alone, as they also depend on the constants $c_{ij}$
in eq.~(\ref{Laguerre_def}). In our case, the Laguerre cells that solve the optimal transport problem\footnote{The weights $c_{ij}$ that define these Laguerre cells depend on the energy fractions $\omega_{i}$ of the particles, that we also call ``weights'', but should not be confused with them.} depend on the particle energy fractions $\omega_i$ via the first constraint in eq.~(\ref{constraints_semi_D}). Enforcing this constraint is precisely how the Laguerre cells are determined in practice.

Once the Laguerre cells are known, the resulting value of the EMD, and hence of the isotropy, is given by
	\begin{equation}
		\label{I_nsph}
		\mathcal{I}_n\left(\left\{p_i\right\}\right)=\frac{1}{4\pi}\sum_{i=1}^{n}\int_{C_i}d\Omega'\,d_i^2(\Omega').
	\end{equation}
The minimisation problem in eq.~(\ref{Insph_def}) has therefore been recast into the problem of determining the Laguerre cells.

\subsection{Two-particle event}
\begin{figure}[t]
\centering
\includegraphics[width=0.50\textwidth]{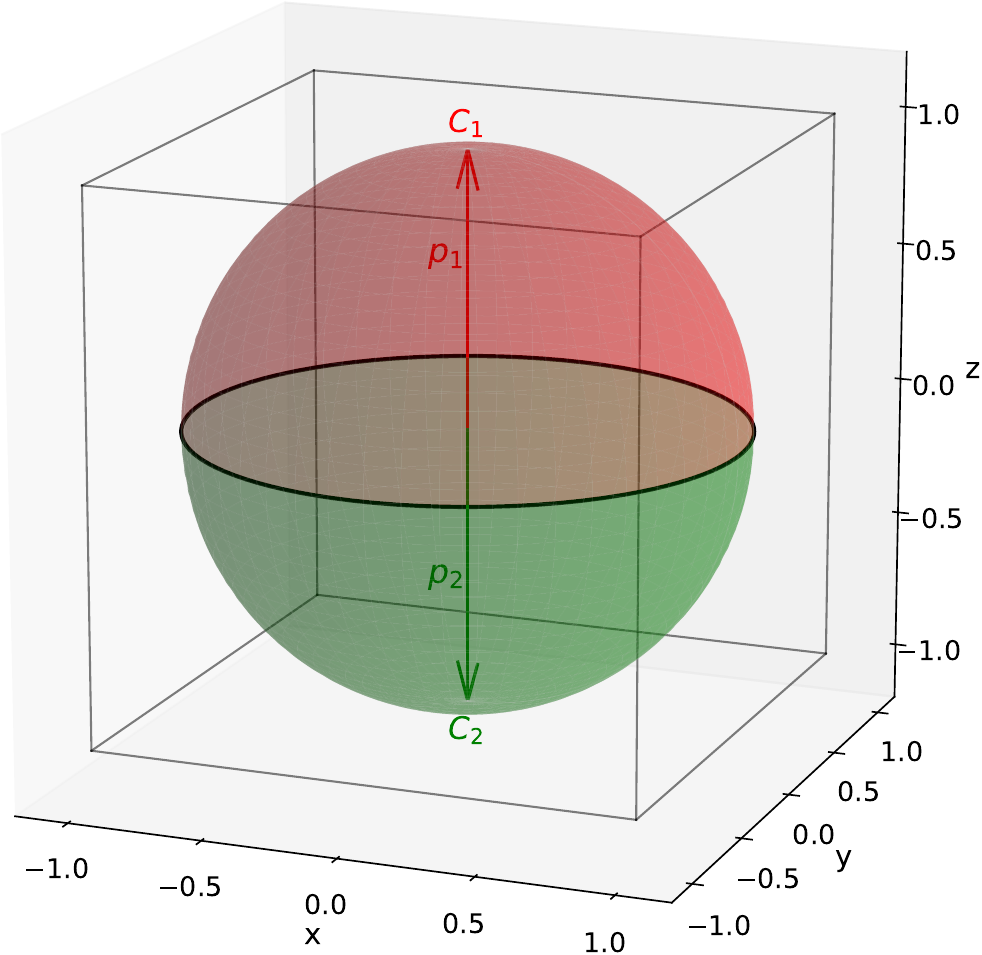}\hfill
\includegraphics[width=0.50\textwidth]{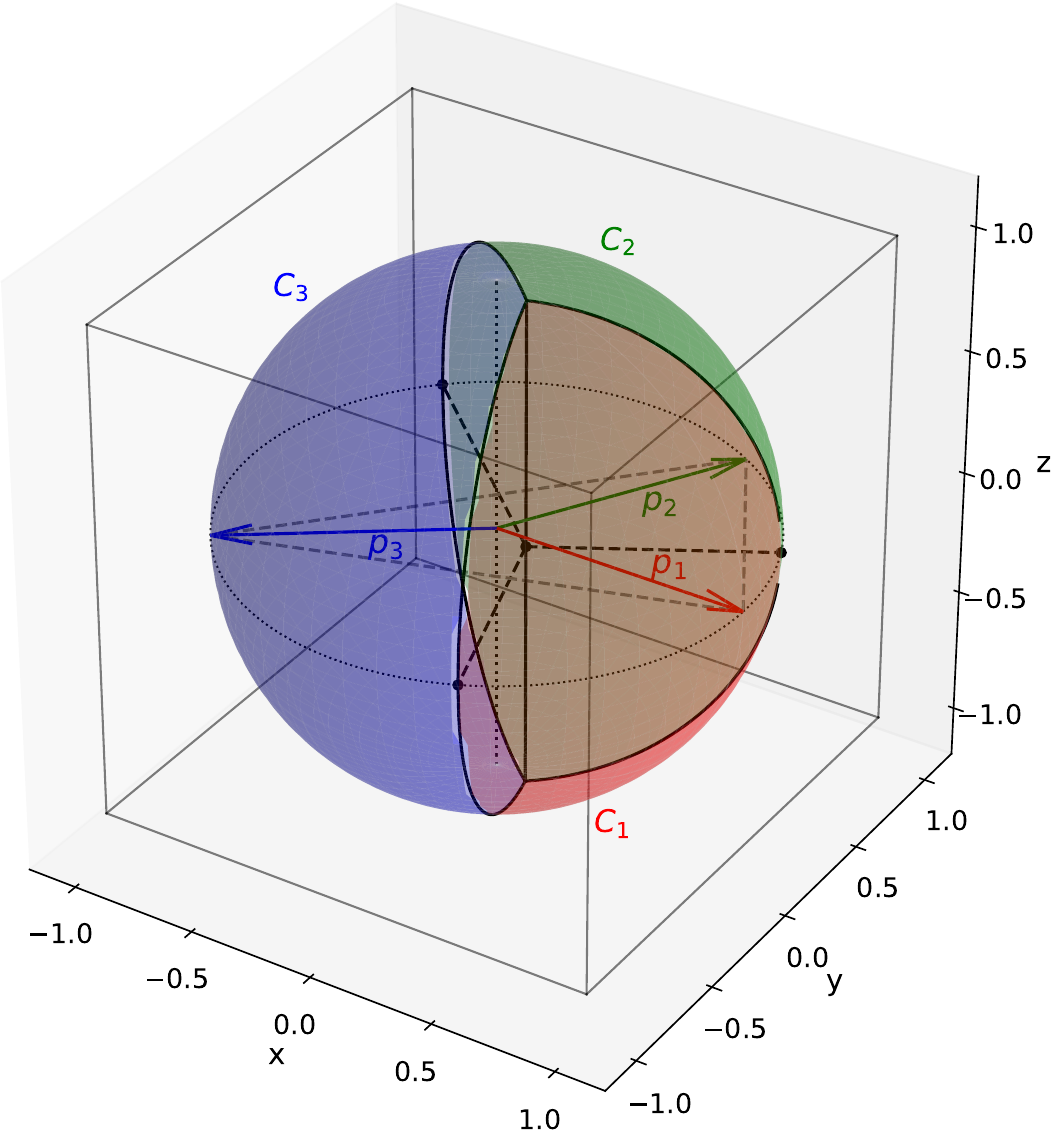}
\caption{On the left, the two-particle case with the division of the sphere into two Laguerre cells, which coincide with Voronoi cells as the two particles carry the same energy; on the right, the division of the sphere into three Laguerre cells for a three-particle configuration, with $x_1=0.45$, $x_2=0.62$ and $x_3=0.93$. Note that the wedges' axis (solid line) does not coincide with the axis of the sphere (dotted line).}
		\label{Laguerre_cells}
\end{figure}

As a first application of the general framework described above, we re-derive the explicit expression for the simplest case, namely the isotropy of a two-particle event~\cite{Cesarotti:2020hwb}. Let us consider two back-to-back particles, each of them carrying half of the total energy. This is the relevant configuration for the process $e^+e^- \to q \bar q$.
Without loss of generality, we impose their spatial momenta to be along the $z$-axis. The Laguerre cells (which, in this case, coincide with Voronoi cells because all energies are equal) are the two hemispheres divided by the plane $z=0$ (see Fig.~\ref{Laguerre_cells}) and the value of the isotropy can be found fully analytically by applying eq.~(\ref{I_nsph}):
	\begin{equation}
		\mathcal{I}_2=2\cdot\frac{1}{4\pi}\int_{-\pi}^{\pi}d\phi'\,\int_{0}^{1}d\cos\theta'\,2(1-\cos\theta')=1.
	\end{equation}
The result is trivial and shows that a two-particle back-to-back event is the most anisotropic one if we work in the c.o.m. frame.

\subsection{Three-particle event}
The case of a three-particle event, which is relevant for $e^+e^- \to q \bar q g$, is more complicated. 
Let us assume that we are in the centre-of-mass frame, so that the total three-momentum vanishes. The final-state particles belong to the same plane: without loss of generality, we choose the plane $z=0$. 
Thanks to azimuthal symmetry, the system has only two degrees of freedom: we choose the energy fractions $x_{1,2}=2E_{1,2}/Q$ of particles 1 and 2 (e.g. the quark and the antiquark). The Laguerre cells that solve the optimal transport problem are spherical wedges with axes perpendicular to the particles' plane, in general different from the sphere axis (see Fig.~\ref{Laguerre_cells}). 
The planes defining the wedges are orthogonal to the lines connecting two particles on the unit sphere. For three particles, this means that the three wedges are fully determined by the position of their intersection vertex on the sphere. This property generalises to larger multiplicities so that for a given $n$ the Laguerre cells are determined by $n-2$ vertices on the unit sphere.
In order to find these Laguerre cells, we impose that their areas be proportional to the energy fractions, according to the first constraint of eq.~(\ref{constraints_semi_D}).
The analytic expressions for the wedges' areas are rather cumbersome and the set of simultaneous equations that enforce the constraints can only be solved numerically. Details are given in appendix~\ref{3particles_computation}.
After finding the cells, we apply eq.~(\ref{I_nsph}) to compute the isotropy:
	\begin{equation}
		\label{I_3sph}
		\mathcal{I}_3\left(x_1,x_2\right)=\frac{1}{4\pi}\sum_{i=1}^{3}\int_{C_i}d\phi'\,d\cos\theta'\,2\left[1-\sin\theta'\cos\left(\phi'-\phi_i\right)\right],
	\end{equation}
where $\phi_i$ is the azimuthal angle of the $i$-th particle of $\mathcal{E}$. As shown in appendix \ref{3particles_computation}, the integration over the Laguerre cells in eq. (\ref{I_3sph}) can be performed analytically. However, the result we obtain for $\mathcal{I}_3\left(x_1,x_2\right)$ is only semi-analytic, due to the numerical step required to determine the Laguerre cells.

\subsection{Numerical checks}
In order to verify our semi-analytic result for $\mathcal{I}_3\left(x_1,x_2\right)$, we check it against two fully numerical approaches. 
The first one is the code from refs.~\cite{Cesarotti:2020hwb,EventIsotropyCode}, which we denote as ``POT''. It works by first discretising the sphere and then solving a fully discrete OT problem. In practice, it uses the HEALPix package~\cite{Gorski_2005} to construct a pseudo-uniform spherical event with a finite number $N$ of particles~\footnote{It is worth pointing out the difference between $n$ and $N$. The parameter $n$ is physical, representing the number of particles of the real event $\mathcal{E}$, whose isotropy is to be computed, whereas $N$ is a non-physical parameter representing the number of particles of the discrete pseudo-spherical event $\mathcal{U}_{N}$. The discretisation of the uniform event $\mathcal{U}$ is needed for the POT numerical evaluation of the isotropy.} and it computes the \EMD~between $\mathcal{E}$ and $\mathcal{U}_{N}$ by using the Python Optimal Transport (POT) library~\cite{Flamary_POT_Python_Optimal}.
The second method we use is a SDOT (Semi-Discrete Optimal Transport) code developed in-house. This code solves directly the semi-discrete problem by identifying the Laguerre cells with an iterative procedure. At each step, it samples the sphere with $M$ points and tests if the constraints in eq.~(\ref{constraints_semi_D}) are satisfied.

\begin{figure}[t]
\centering
\includegraphics[width=\textwidth]{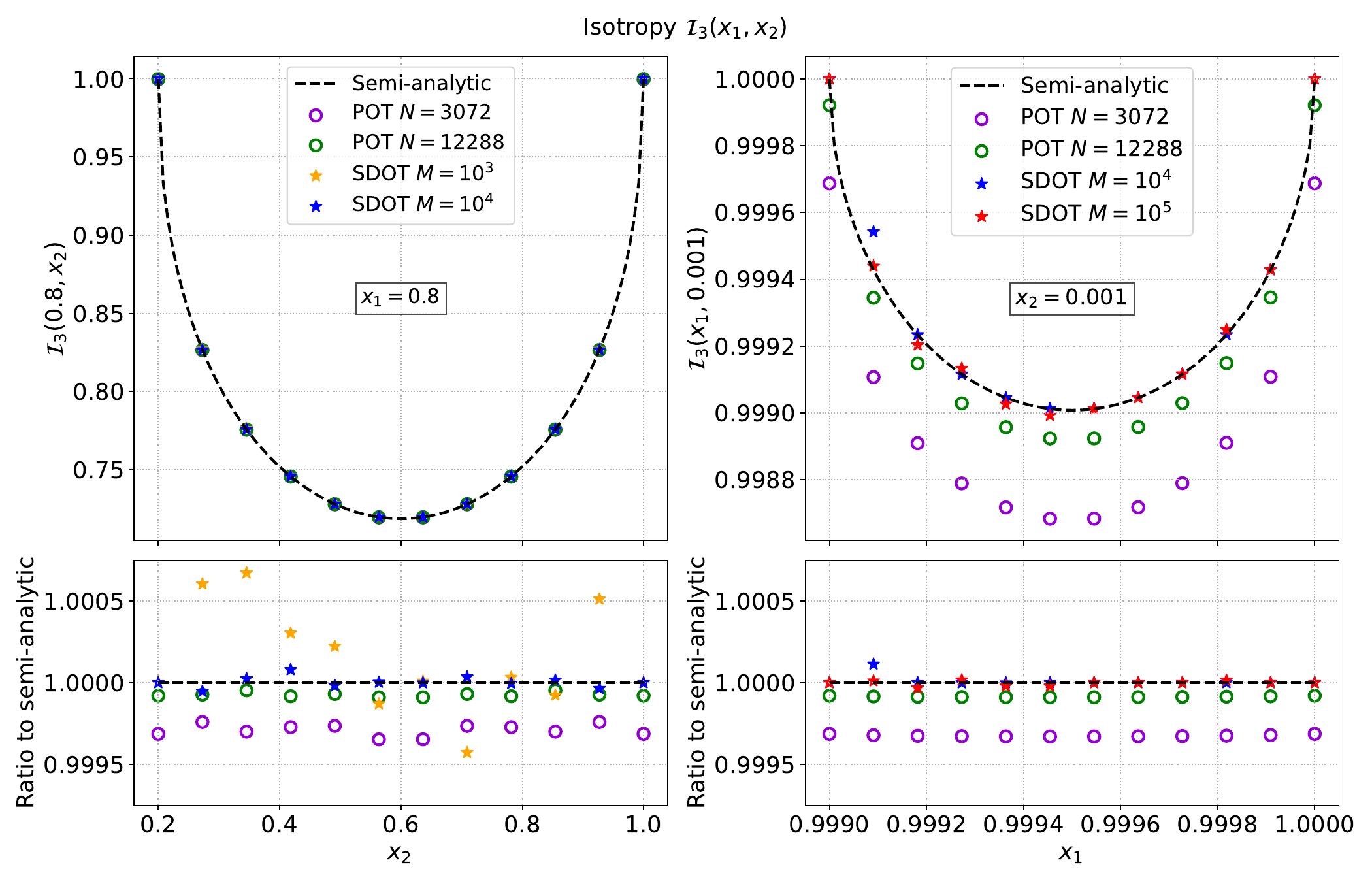}
\caption{Event isotropy of a three-particle final state in spherical geometry: comparison between the semi-analytic computation and the fully numerical POT and SDOT results. On the left, the energy fraction $x_1$ is fixed, while $x_2$ takes values in the interval $\left[1-x_1,1\right]$: the line represents the semi-analytic result, the points represent the POT result and the stars the SDOT result; on the right, $x_2$ is fixed, while $x_1$ varies in the interval $\left[1-x_2,1\right]$.}
\label{isotropy3_plot_3sph}
\end{figure}

Fig. \ref{isotropy3_plot_3sph} shows the value of the isotropy as a function of the energy fractions $x_1$ and $x_2$. The left plot presents results for a configuration where all three particles have comparable energies, while the right plot addresses a situation where one of the particles is very soft and almost collinear.
We can see that the agreement of our semi-analytic calculation with the fully numerical ones increases as $N$ and $M$ increase, as expected. We also note that, while not yet fully visible in these plots, the performance of the POT and SDOT numerical approaches degrades in the soft-collinear configuration, again as expected. 

We close this section by stressing the advantages of a semi-analytic expression for event isotropy, albeit for a low final-state multiplicity, when performing perturbative calculations. It allows one to avoid discretisation errors that are inherent to numerical evaluations, and hence to obtain reliable results also in the soft and collinear regions of the phase space. We will come back to this point in the next section.

\section{Isotropy distribution in perturbative QCD}\label{sec:QCD}

Thus far, we have discussed only the mathematical approach to the computation of event isotropy. Now, we move our discussion to physics in order to exploit perturbative QCD and obtain a theoretical description of the distribution for event isotropy in $e^+e^-$ collisions at energy $Q=M_Z$.
Following what is usually done in the case of thrust, we find it convenient to present results for the differential distribution of the observable
\begin{equation}
    \label{iota}
    \iota\equiv1-\mathcal{I}.
\end{equation}
This way, $\iota=0$ (i.e.\ a maximally anisotropic configuration) at Born level.
\subsection{Fixed-order calculation}\label{sec:FO}
In order to have a nonzero value for $\iota$ we need at least one emission. Therefore, the leading-order distribution in QCD is obtained by integrating the $\ord{\as}$ matrix element for the $e^+e^- \to \gamma^* \to q \bar q g$ final state
\begin{equation}
    \label{eq:matrix_element}
    \frac{d\sigma}{dx_1dx_2}=
   \sigma_0  \frac{\as \cf}{2\pi}  \frac{x_1^2 + x_2^2}{(1-x_1)(1-x_2)} \, ,
   \qquad\qquad 
   \sigma_0 = \frac{4 \pi \alpha_{em}^2}{3 Q^2} 3 \sum^{\nf} e_q^2 \, ,
\end{equation}
against the definition of the observable. For this calculation, we can exploit the three-particle semi-analytic result obtained in the previous section. We have then 
\begin{equation}
		\label{dsigma_qbarqg_diota}
		\frac{1}{\sigma_0}\frac{d\sigma}{d\iota}=\frac{1}{\sigma_0}\int{}dx_1dx_2\,
        \frac{d\sigma}{dx_1dx_2}\,
        \delta\left(\iota-\iota_3\left(x_1,x_2\right)\right) \, ,
	\end{equation}
where $\iota_3\left(x_1,x_2\right)= 1- {\cal I}_3\left(x_1,x_2\right)$ and ${\cal I}_3$ is given in eq. (\ref{I_3sph}).

\begin{figure}[t]
\centering
\includegraphics[width=\textwidth]{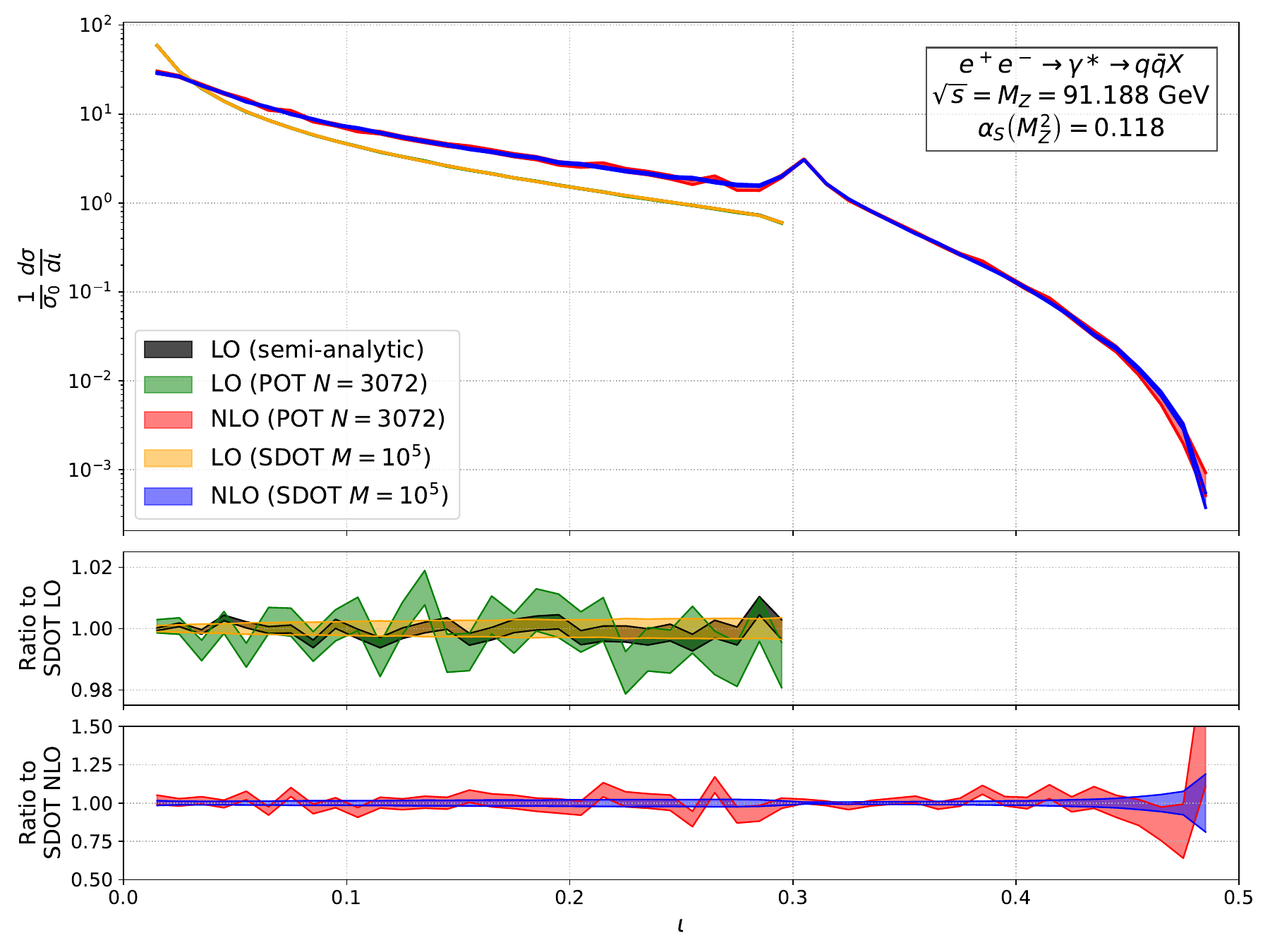}
\caption{\label{fig:iota_distribution} Fixed-order differential distribution of $\iota$. The POT histogram is obtained by generating $5\cdot10^5$ events with \eventtwo~and computing their isotropy using $N=3072$; the SDOT histogram is obtained with $3\cdot10^7$ \eventtwo~events using $M=10^5$; the semi-analytic result is obtained integrating numerically our semi-analytic formula with the $q\bar q g$ matrix element using $5\cdot10^9$ events. Statistical uncertainties are included.}
\end{figure}

In Fig.~\ref{fig:iota_distribution}, we compare the $\ord{\as}$ distribution from this semi-analytic approach with the one obtained  generating events with \eventtwo\ \cite{Catani:1996jh,Catani:1996vz} and computing their isotropy with both the POT and SDOT codes.
We observe that the three $\ord{\as}$ distributions are in agreement within uncertainties.
The plot also shows the $\ord{\as^2}$ fixed-order distribution.
At this order, we do not have an in-house matrix-element generator and, further, we do not have a semi-analytic result for $\iota$ in the four-parton case. 
Thus, we rely on \eventtwo\ and use POT and SDOT to compute $\iota$. Again, we find that they are in agreement within statistical uncertainties.

Comparing the $\ord{\as}$ and the $\ord{\as^2}$ curves we see that large perturbative corrections appear in two distinct regions.
At small $\iota$, events are pencil-like and anisotropic and the distribution is dominated by soft and collinear emissions. Perturbative coefficients are affected by large logarithmic corrections that need to be resummed in order to obtain reliable predictions. This issue will be addressed in the next section.
At the other end of the spectrum, i.e.\ large $\iota$, the isotropy distribution reaches a maximum, which represents the most isotropic configuration for a set of $n$ particles. At $\ord{\as}$, where we have a three-particle final state, the maximum is reached when the three particles carry the same amount of energy ($Q/3$) and their spatial momenta point to the vertices of an equilateral triangle: this value can be  computed exactly and it is equal to $\iota_3^\text{max} = 3\sqrt{3}/4-1 \simeq 0.299038$ (details in appendix~\ref{fully-analytical}).
At $\ord{\as^2}$ the endpoint of the distribution is larger: a tetrahedral configuration with equal energies, the most isotropic configuration with $n=4$, gives $\iota_4^\text{max}  = 2\sqrt{6}\arctan(\sqrt{2})/\pi-1   \simeq 0.489715$ (see again appendix~\ref{fully-analytical}).~\footnote{The maximum value $\iota_n^\text{max}$ for a generic $n$-particle final state increases as $n$ increases, and it approaches one when $n\rightarrow+\infty$: the ideal isotropic configuration can be reached only with an infinite number of particles, otherwise there will always be some privileged directions.}
The opening up of the phase space explains the sizeable perturbative corrections at large $\iota$ when going from $\mathcal{O}\left(\alpha_S\right)$ to $\mathcal{O}\left(\alpha_S^2\right)$.

\subsection{Resummation}\label{sec:res}
We now want to study the behaviour of the isotropy distribution in the small-$\iota$ region, i.e. $\mathcal{I}\to 1$. In this regime, fixed-order perturbation theory becomes unreliable because large logarithmic corrections are generated by the emission of soft and collinear partons and they must be resummed to all orders. 
We aim to determine this resummation at next-to-leading logarithmic (NLL) accuracy.

In order to understand the NLL structure of the isotropy distribution, following e.g.~\cite{Banfi:2004yd}, we must first determine the behaviour of the observable in the presence of a soft and collinear emission. 
We already know that at Born level, with just two partons $\bar 1$ and $\bar 2$,
respectively at $\theta_1=\pi/2$, $\phi_1=0$ and $\theta_2=\pi/2$, $\phi_2=\pi$,
we have $\iota=0$ and the two Laguerre cells are just the hemispheres: $C_{\bar 1}= \Sigma_R$ and $C_{\bar 2}= \Sigma_L$, with $\Sigma_R=\left\{(\theta',\phi'):\phi'\in(-\pi/2,\pi/2]\right\}$ and $\Sigma_L=\left\{(\theta',\phi'):\phi'\in(-\pi,-\pi/2]\cup(\pi/2,\pi]\right\}$, where the azimuthal angle $\phi'\in(-\pi,\pi]$.

Let us add a soft and collinear emission: without loss of generality, we consider a third particle emitted from particle 2 and introduce the variables
\begin{equation}
    \label{z_theta_def}
    z\equiv\frac{x_3}{x_2+x_3},\hspace{10ex}\theta\equiv\phi_3-\phi_2\pmod{2\pi}. 
\end{equation}
Thus, $z$ is the fraction of energy of the emitted particle with respect to the energy of the emitting leg and $\theta$ is the angle between the two particles after the splitting. The soft and collinear condition is $\theta,z\ll1$. In this configuration, the kinematics of the system is
\begin{subequations} \label{soft-momenta}
\begin{align}
    & x_1=1-\frac{1}{4}z\theta^2+\mathcal{O}\left(z^2\theta^2\right), & \phi_1&=0,
    \\
    & x_2=1-z+\frac{1}{4}z\theta^2+\mathcal{O}\left(z^2\theta^2\right), & \phi_2&=\pi-z\theta+\mathcal{O}\left(z\theta^3\right),
    \\
    & x_3=z+\mathcal{O}\left(z^2\theta^2\right), & \phi_3&=-\pi+(1-z)\theta+\mathcal{O}\left(z\theta^3\right).
\end{align}
\end{subequations}
The computation of isotropy starts from eq.~(\ref{I_nsph}) where  $\mathcal{I}_3(z,\theta)$ is written as the sum over three Laguerre cells $C_{1,2,3}$ corresponding to the three particles in the event. 
We first add and subtract the isotropy $\mathcal{I}_2=1$ from the Born event with particles $\bar 1$ and $\bar 2$. 
After some relatively lengthy but simple manipulations (e.g\ exploiting the fact that $d_1\left(\Omega'\right)=d_{\bar{1}}\left(\Omega'\right)$), we arrive at the following expression:
\begin{align}
    \label{is3-sc-cntd}
    \mathcal{I}_3(z,\theta) & = 1+\frac{1}{4\pi}\int_{C_3}d\Omega'\,\left(d_3^2\left(\Omega'\right)-d_2^2\left(\Omega'\right)\right)
    +\frac{1}{4\pi}\int_{\Sigma_L}d\Omega'\,\left(d_2^2\left(\Omega'\right)-d_{\bar{2}}^2\left(\Omega'\right)\right) \nonumber
    \\
    & \phantom{=} +\frac{1}{4\pi}\int_{C_1\setminus\Sigma_R}d\Omega'\,\left(d_1^2\left(\Omega'\right)-d_2^2\left(\Omega'\right)\right).
\end{align}
The first two integrals that appear in eq.~(\ref{is3-sc-cntd}) represent, respectively, the genuine contribution to the isotropy due to the additional soft and collinear emission $3$, and the contribution due to recoil of the hard parton $\bar 2 \to 2$, respectively, while we cannot attribute an intuitive description to the third term. 

The evaluation of eq.~(\ref{is3-sc-cntd}) in the soft-collinear limit then follows from a few geometrical observations.
Indeed, following the construction of the Laguerre cells, one quickly realises that in the soft-collinear limit $C_3$ is a very small region (of area  $2\pi x_3\approx 2\pi z\ll 1$) located around $\theta=\pi/2$ and $\phi=-\pi/2$.
This is actually sufficient to find (again after some algebra)
\begin{subequations}\label{recoil}
\begin{align}
\int_{C_3}d\Omega'\,\left(d_3^2\left(\Omega'\right)-d_2^2\left(\Omega'\right)\right) & = -4\pi{}z\theta+\mathcal{O}\left(z^3\theta\right)+\mathcal{O}\left(z^2\theta^2\right),
\\
\int_{\Sigma_L}d\Omega'\,\left(d_2^2\left(\Omega'\right)-d_{\bar{2}}^2\left(\Omega'\right)\right) & = \mathcal{O}\left(z^2\theta^2\right), \\
\int_{C_1\setminus\Sigma_R}d\Omega'\,\left(d_1^2\left(\Omega'\right)-d_{2}^2\left(\Omega'\right)\right) & = \mathcal{O}\left(z^2\theta^3\right).
\end{align}
\end{subequations}
Thus, we find the event isotropy in the presence of a soft and collinear emission to be 
\begin{equation}\label{is3-sc-final}
    \mathcal{I}^\text{sc}_3(z,\theta)= 1- z \theta \quad \Rightarrow \quad \iota_3^\text{sc} \equiv 1- \mathcal{I}^\text{sc}_3 = z \theta.
\end{equation}
The results in eqs.~(\ref{recoil}) tell us that event isotropy is recoil free at NLL.
This information is sufficient to achieve the following NLL resummed cumulative distribution, see e.g.~\cite{Catani:1992ua,Banfi:2004yd, Marzani:2019hun}~\footnote{Because we also include the constant $C_1$, the accuracy we reach is often referred to in the literature as NLL$^\prime$ accuracy.}
\begin{equation}\label{eq:cumulative-nll}
\Sigma_{\text{res}} (\iota)= \left( 1+ \frac{\as \cf}{2 \pi} C_1\right)   \mathcal{F}(R')\, e^{-R(\iota)}, 
\end{equation}
where $R'= -\frac{d R}{d \log \iota}$.
Having determined the behaviour of the observable in the soft and collinear limit eq.~(\ref{is3-sc-final}), the calculation of the radiator function $R$ at NLL can be performed with standard techniques, as summarised in appendix~\ref{app:radiator}. 
Furthermore, in appendix~\ref{app:C1} we describe the calculation of the resummation constant $C_1$.

The function $\mathcal{F}$ is the so-called multiple-emission contribution.
For additive observables~\footnote{We say that an observable is additive if, in the presence of many soft emissions that are well separated in rapidity but all contribute similarly to the value of the observable $v$, we can write $v = \sum_{i=2}^n v_i$, where $v_i$ are the contributions of each emission. Thrust, jet mass, and jet angularities are examples of additive observables.} this function is known in closed form~\cite{Catani:1992ua}:
\begin{equation}
    \mathcal{F}(R') = \frac{e^{-\gamma_E R_{LL}'}}{\Gamma(1 + R_{LL}')},
\end{equation}
where $R_{LL}$ is the LL part of the radiator. 
The fact that event isotropy is an additive observable follows again from a geometric argument. Let us assume that we have a series of soft-collinear emissions. Each of these emissions will be associated with its own Laguerre cell. If, as above, we take a Born event along the $x$ axis, the same arguments as for a single emission imply that each Laguerre cells associated with each of the soft-collinear emissions will be a small region (of area parametrically suppressed by the momentum fraction of the emissions) around the $y-z$ plane of the event. 
One could then construct an expression similar to eq.~(\ref{is3-sc-cntd}) where the first integral (over $C_3$) would be replaced by a sum over the full set of soft-collinear emissions and the remaining terms would only give subleading corrections.
In this setup, isotropy for a set of soft-collinear emissions would indeed be a simple sum of independent contributions for each emission and hence be additive.
In appendix~\ref{app:ME} we formalise the above argument for the additivity of isotropy in the case of two soft and collinear emissions. Our findings are supported by a numerical study in the same limit. We believe that our analytic argument generalises to the case of any number of emissions, thus justifying the use of the above expression for $\mathcal{F}$.

To test our all-order calculation, we compute the expansion of the resummation at first and second order in the strong coupling.
The expansion of the cumulative reads
\begin{align}\label{eq:expansion}
    \Sigma_{\text{res}} (\iota)&= \Bigg[1+ \frac{\as\cf }{2\pi}\left(c_{12}\log^2 \iota +c_{11}\log\iota +c_{10} \right)\nonumber\\&+
    \left(\frac{\as\cf }{2\pi}\right)^2\left(c_{24}\log^4 \iota +c_{23}\log^3 \iota+c_{22}\log^2 \iota + c_{21}\log \iota \right)  \Bigg] +\ord{\as^3},
\end{align}
where $\as=\as(\mur^2)$.
Our resummed calculation is expected to fully reproduce all the above $c_{1j}$ coefficients, as well as $c_{2j}$, with $j\ge 2$. 
By expanding the resummed expression of app.~\ref{resum_details}, we find:
\begin{subequations}
\label{eq:coeffasas2}
    \begin{align}
c_{12}&=-4, \qquad\qquad c_{11}=8 \log2 -6, \qquad\qquad c_{10}=C_1,\\
    c_{24}&=8, \qquad\qquad\quad c_{23}=-4\left( 8\log{2}-6-\frac{8\pi\beta_0}{3C_F}\right), \\
c_{22}&=\frac{1}{2}\left(8\log{2}-6\right)^2 
-\frac{16}{3}\pi^2-4 C_1
-\frac{4\pi\beta_0\left(8\log{2}-3+ 2 \log\frac{\mur^2}{Q^2}\right)+4 K}{C_F},
\end{align}
\end{subequations}
where $\beta_0$, $K$, and $C_1$ are given in eqs.~(\ref{eq:beta-function}), (\ref{eq:cmw-K}), and~(\ref{eq:C1-result}), respectively. 
\footnote{For the coefficient $c_{21}$ we find
\begin{equation}
    c_{21} =(8 \log 2 -6)C_1.
\end{equation}
However, $c_{21}$ is a NNLL contribution and therefore its complete determination goes beyond the accuracy of our resummation, receiving additional contributions starting at NNLL.}
The differential distribution is obtained from eq.~(\ref{eq:cumulative-nll}) by differentiation:
\begin{equation}\label{eq:diff-nll}
    \frac{1}{\sigma_0}\frac{ d \sigma_{\text{res}} }{d \iota}=  \frac{d \Sigma_{\text{res}} (\iota)}{d \iota},
\end{equation}
while its expansion can be straightforwardly obtained by differentiating eq.~(\ref{eq:expansion}).

\begin{figure}
    \centering
\includegraphics[width=\textwidth]{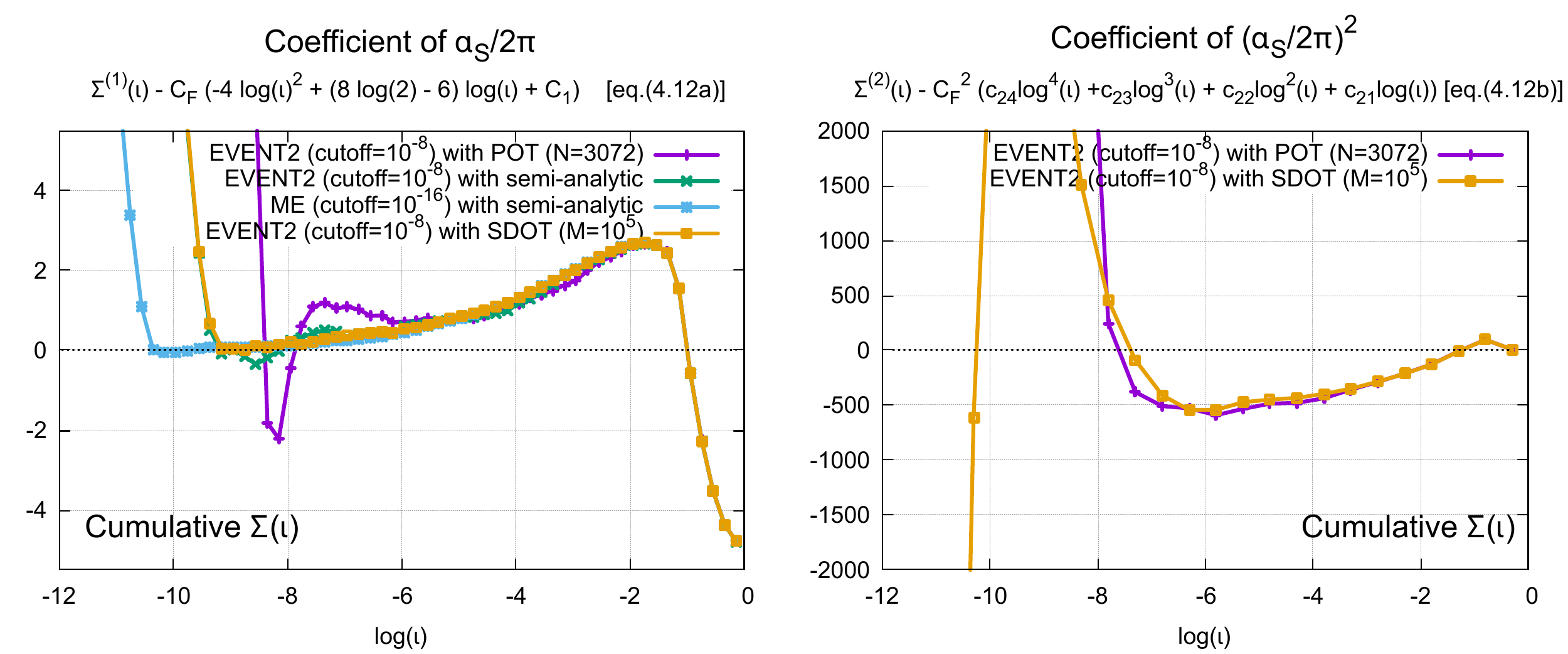}
\includegraphics[width=\textwidth]{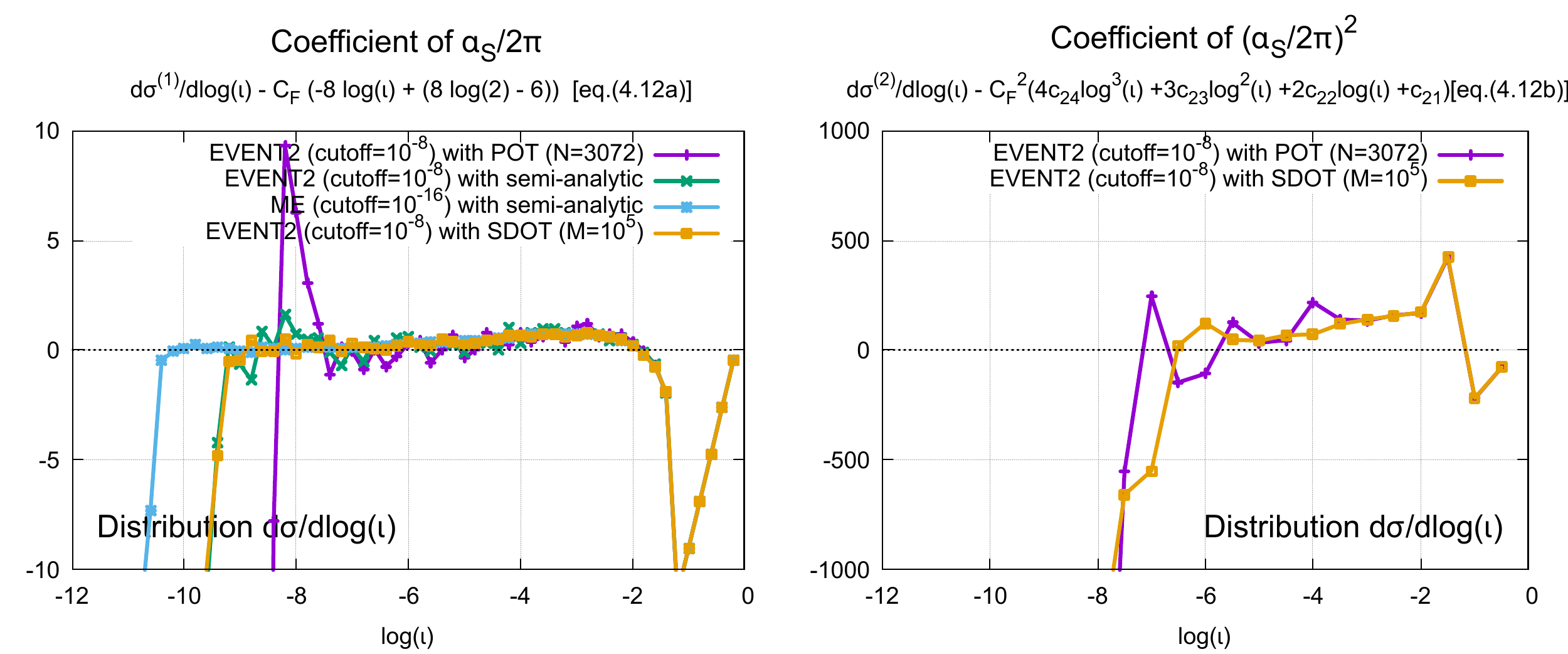}
	\caption{Fixed-order checks of the expansion of the resummation to $\ord{\as}$, on the left, and to $\ord{\as^2}$, on the right. We plot the difference between the coefficients extracted from \eventtwo\  and the theoretical predictions in eqs.~(\ref{eq:coeffasas2}). Plots at the top are for the cumulative distribution $\Sigma$, the bottom ones for the differential distribution. See footnote~\ref{foot:c2} about the calculation of the ${\cal O}(\as^2)$ coefficient of the cumulative. The same expedient has been used for the ${\cal O}(\as)$ coefficient of the differential distribution calculated with \texttt{ME} (see footnote \ref{foot:me}), using of course this time the value of the ${\cal O}(\as)$ term in eq.~(\ref{eq:totalxsect}).}
		\label{fig:smalliota}
\end{figure}

In order to check these results, we compare them to the ${\cal O}(\as)$ and ${\cal O}(\as^2)$ coefficients of the distribution $d\sigma/d\log\iota$ and of the cumulative $\Sigma(\iota)$ that can be extracted from the \eventtwo~code.\footnote{
\label{foot:c2}\eventtwo~does not contain the virtual term at order $\as^2$, when calculating the coefficient of the cumulative at this order. While technically it is not needed for the comparison that we perform here, we include it nonetheless, and we extract it using the known analytic result for the inclusive cross-section into hadrons:
\begin{equation}
\label{eq:totalxsect}
\sigma = \sigma_0 \left(1 + \frac{\as}{2 \pi} \cf \frac{3}{2} +  \left(\frac{\as}{2 \pi}\right)^2 c_2\right),
\end{equation}
with~\cite{Baikov:2012zn}
\begin{equation}
c_2 = 4\,(1.9857 - 0.1152\, \nf).
\end{equation}
}
This is done in Fig.~\ref{fig:smalliota}, where we show the difference between the fixed-order distribution and the expansion of the resummation.
The plots at the top are for the cumulative distribution $\Sigma$, while the bottom ones are for the differential distribution. On the left, we show the  $\ord{\as}$ checks, while on the right, the $\ord{\as^2}$ ones.
At $\ord{\as}$, we control all logarithms and the constant contribution $c_{10}$ in the cumulative distribution. Thus, we expect both differences computed at $\ord{\as}$ to vanish at small $\iota$. This is confirmed by both plots on the left-hand side. 
It is worth noting that, in order to capture the asymptotic behaviour, we have to reach very small values of $\iota$. This results in numerical instabilities when we use the POT code to evaluate the observable. The SDOT code allows us to go deeper in the infrared, while using \eventtwo. However, it is only when using an in-house implementation, denoted \texttt{ME}~\footnote{\label{foot:me}\texttt{ME} is a Monte Carlo calculation of the isotropy distribution which uses eq. (\ref{eq:matrix_element}) for the ${\cal O}(\as)$ cross section.
} in the plots, with the semi-analytic evaluation of isotropy that we can meaningfully probe the asymptotic behaviour at small $\iota$.

With the above considerations in mind, we move to the $\ord{\as^2}$ checks. Here we have less leverage because we have to rely on \eventtwo~simulations with the isotropy computed with POT or SDOT. 
In this case, we expect the difference between the cumulative and the expansion of the resummation (top right plot in Fig.~\ref{fig:smalliota}) to behave as a straight line, while for the differential distribution (bottom right plot) we expect a constant at small $\iota$. The behaviour we observe in the plots is broadly compatible with such expectations.

\subsection{Matching}
In order to obtain a reliable perturbative prediction across the $\iota$ spectrum, we match the NLL resummation to the fixed-order calculation. We use multiplicative matching, which for the cumulative distribution reads
\begin{equation}\label{eq:matching}
    \Sigma_{\text{matched}} = \Sigma_{\text{res}} \left\{1 + \frac{\as}{2 \pi} \left(\Sigma^{(1)}- \Sigma^{(1)}_{\text{res}} \right)+\left(\frac{\as}{2 \pi} \right)^2\left[\left(\Sigma^{(2)}- \Sigma^{(2)}_{\text{res}} \right) -
    \Sigma^{(1)}_{\text{res}}\left(
    \Sigma^{(1)}- \Sigma^{(1)}_{\text{res}}\right)\right]  \right\}, 
\end{equation}
where $\Sigma^{(i)}_{\text{res}}$ corresponds to the coefficient of order $i$ in the expansion of the resummed expression.
In order to force the resummation to have the same kinematic endpoint $\iota_i^\text{max}$ as the fixed order it is matched to, we also shift
the relevant logarithms by adding power-suppressed contributions:
\begin{equation}\label{eq:log_mod}
    \log \frac{1}{\iota}\to \log \left(1+\frac{1}{\iota}-\frac{1}{\iota_i^\text{max}}\right),
\end{equation}
where $i=3,4$ for LO and NLO matching, respectively.

\begin{figure}
\centering
\includegraphics[width=\textwidth]{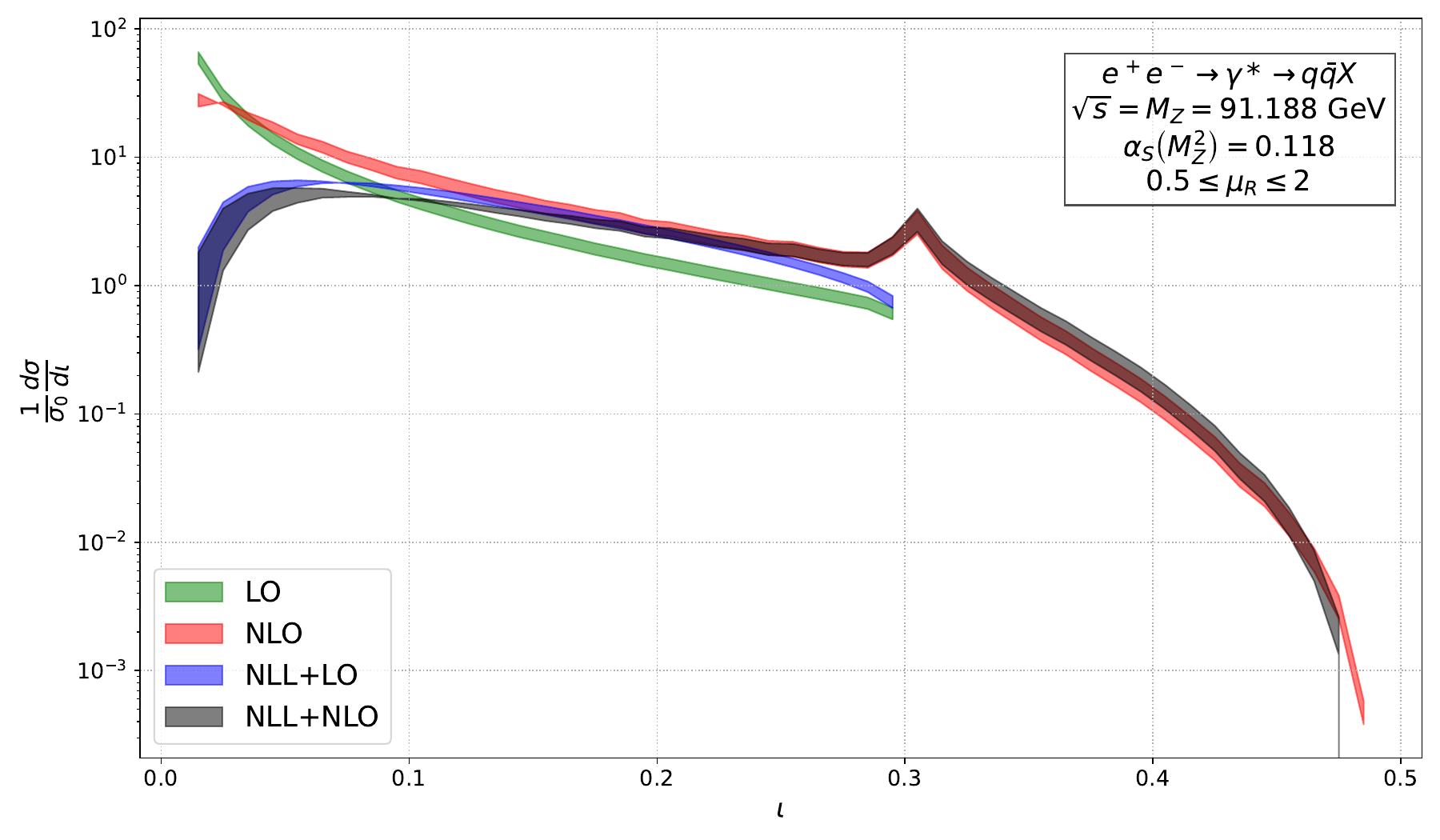}
\caption{Comparison of the fixed-order differential distribution to the resummed and matched one, at different perturbative accuracies, as indicated in the plot legend. The bands indicate the theoretical uncertainty and are obtained by varying the renormalisation scale $\mur$.}\label{fig:matching}
\end{figure}
Our final results for the NLO+NLL differential distribution, which is obtained by differentiating eq.~(\ref{eq:matching}), are collected in Fig.~\ref{fig:matching}. We show four curves: the fixed-order distributions, at LO and NLO, i.e.\ the same as Fig.~\ref{fig:iota_distribution}, and the matched distributions at NLL+LO and NLL+NLO.  In order to estimate the perturbative uncertainty, we vary the renormalisation scale $\mur$ around the hard scale $Q$ by a factor of 2.~\footnote{For phenomenological studies, it would also be important to assess the uncertainty associated with the accuracy of the resummation. This is usually done by rescaling the argument of the logarithms being resummed by a factor $x_Q$, which is then varied around unity. This procedure does not affect the fixed-order result and is not considered in the present proof-of-principle study.}

As already commented, the FO predictions are not reliable at small $\iota$ because of the presence of large logarithmic corrections. Indeed the behaviour of the matched results is remarkably different compared to the FO, as expected. 
At intermediate values of the observable, i.e.\ away from the endpoints, say $0.15<\iota<0.25$, we would expect FO perturbation theory to be reliable. However, we obtain large corrections in going from LO to NLO. Interestingly the resummation appears to capture the bulk of these corrections. Indeed the NLO, NLL+LO and NLL+NLO curves are all very similar in this region, with only the pure LO being substantially lower. We have checked that this effect is in part due to the presence of the resummation constant $C_1$. 
The large-$\iota$ region is instead characterised by many emissions giving rise to isotropic events. Here, both the approaches considered in this paper are expected to fail, either because we have a limited number of emissions (FO) or because such emissions are soft and/or collinear (resummation). 
It would be very interesting to explore whether non-perturbative methods, such as the ones discussed for instance in~\cite{Cesarotti:2020uod}, can give us a handle to describe this part of the distribution.

\section{Conclusion and Outlook}\label{sec:conclusions}
In this work we have performed a first-principle perturbative calculation of an observable defined through the Energy Mover's Distance (\EMD), namely event isotropy ($\mathcal{I}$) in $e^+e^-$ collisions. The aim was to place this geometrically motivated observable on a similar theoretical footing as more traditional event shapes, and to clarify how its definition in terms of optimal transport can be reconciled with standard tools of perturbative QCD, such as fixed-order calculations and resummation.

A key ingredient of our analysis is a semi-analytic treatment of the optimal transport problem that defines event isotropy. 
This makes the observable amenable to analytic calculations while preserving its interpretation as a distance in the space of energy flows. On this basis, we have  derived perturbative predictions for the event-isotropy distribution, combining fixed-order results at NLO in QCD, with the NLL resummation of logarithmically enhanced contributions in the region $\mathcal{I}\to 1$, i.e.\ the anisotropic limit, which is dominated by soft and collinear radiation.
Thus, our final matched result achieves NLL+NLO accuracy.\footnote{Event isotropy is expected to be affected by non-perturbative corrections. Since this paper focuses on a pioneering calculation of isotropy in perturbative QCD, we defer the investigation of non-perturbative effects to future work. }

We envisage a few natural extensions of this work. A first important step is the generalisation of our framework to proton--proton collisions. In that case, the presence of initial-state radiation, the underlying event, and pileup introduce additional sources of soft activity that are expected to have a non-trivial interplay with the global nature of the \EMD. In this context, we plan to extend our calculation to the ring geometry, more natural for $pp$ collisions, and to perform a comparison with experimental data~\cite{ATLAS:2023mny}.

From a more conceptual point of view, event isotropy is qualitatively different from other observables that measure the distance to an idealised configuration with a finite number of hard directions, e.g.\ thrust or $N$-(sub)jettiness~\cite{Stewart:2010tn,Thaler:2010tr}, even if they can also be expressed in terms of EMD. In those cases, perturbation theory provides a natural starting point, as one can generate configurations with a fixed number of energetic prongs dressed by soft and collinear radiation. By contrast, a truly uniform energy distribution lies beyond the direct reach of perturbation theory. This raises the question of whether perturbative methods alone are sufficient to describe the data, in particular in the fully isotropic limit $\mathcal{I}\to 0$, or whether one should explore non-perturbative descriptions of event isotropy~\cite{Cesarotti:2020uod}.

\paragraph{Acknowledgements.} 
We thank Cari Cesarotti and Jesse Thaler for inspiring discussions and comments.
S.M. wishes to thank Simone Di Marino for useful discussions about mathematical aspects of optimal transport. 
M.C. thanks Gabriel Peyr\'e and Bruno Levy for conversations and help with semi-discrete optimal transport codes.
S.M. wishes to thank LPTHE for hospitality during the course of this work.

\FloatBarrier
\appendix 

\section{Details of the three-particle computation}
	\label{3particles_computation}

In this appendix, we present the details of the calculation of event isotropy for a three-particle final state. 

\subsection{Partition into three Laguerre cells}
We represent the particles' directions as unit vectors in the plane $z=0$. Furthermore, exploiting the azimuthal symmetry of the problem, we fix the direction of particle 1 and write the expressions of the four-momenta of the particles as
	\begin{equation}
		\label{four_momenta}
		p_1=\frac{x_1Q}{2}\left(1,1,0,0\right),\quad p_2=\frac{x_2Q}{2}\left(1,\cos\phi_2,\sin\phi_2,0\right), \quad p_3=\frac{x_3Q}{2}\left(1,\cos\phi_3,\sin\phi_3,0\right),
	\end{equation}
with $\phi_2\in[0,\pi]$ and $\phi_3\in[-\pi,0]$.

\begin{figure}
    \centering
\includegraphics[width=0.495\textwidth]{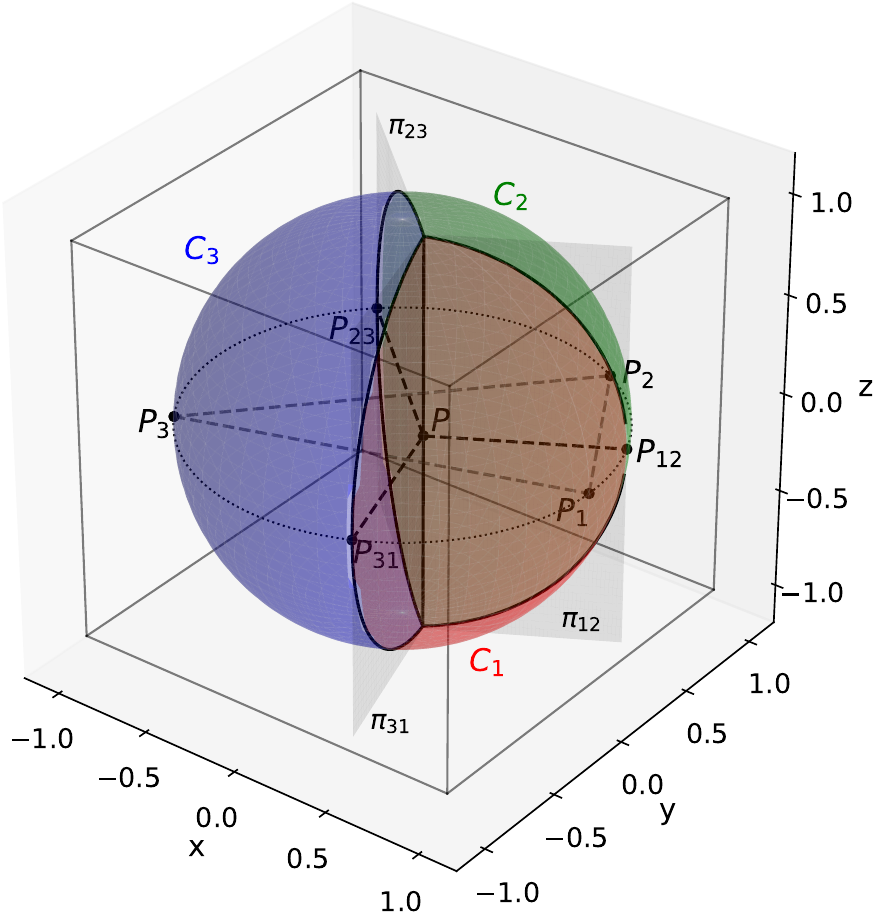}
\includegraphics[width=0.495\textwidth]{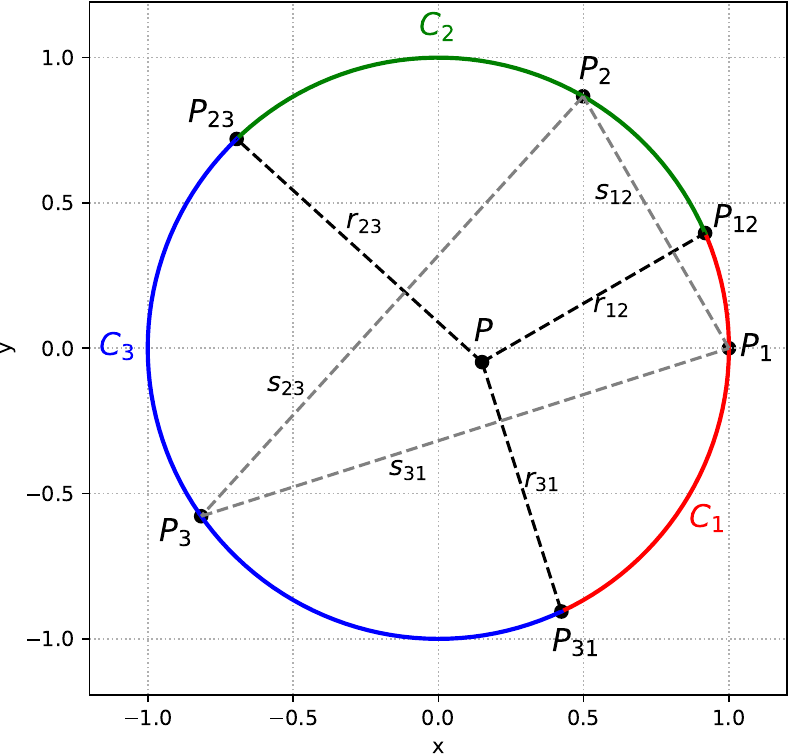}
	\caption{On the left, graphical visualisation of three Laguerre cells on the sphere with three particles: $x_1=0.45$, $x_2=0.62$ and $x_3=0.93$; on the right, the projection on the $z=0$ plane.}
		\label{Laguerre_cells_3sph}
	\end{figure}

On the unit circle in the $z=0$ plane, we consider the endpoints $P_1\left(1,0\right)$, $P_2\left(\cos\phi_2,\sin\phi_2\right)$, and $P_3\left(\cos\phi_3,\sin\phi_3\right)$ of the particle directions, and the segments $s_{ij}$ (with $ij\in\{12,23,31\}$) connecting them, see Fig.~\ref{Laguerre_cells_3sph}. Let $P\left(r\cos\psi,r\sin\psi\right)$ be a point inside the circle. From $P$, we draw the normals $r_{ij}$ to the segments $s_{ij}$: these lines will intersect the circumference in three points that we denote by $P_{ij}\left(\cos\xi_{ij},\sin\xi_{ij}\right)$. The angles $\xi_{ij}$ are related to the position of the point $P$ through the relation 
\begin{equation}
		\label{varphi_ij}
		\tan\frac{\xi_{ij}}{2}=\frac{\sin\phi_j-\sin\phi_i\pm\sqrt{2\left[1-\cos\left(\phi_j-\phi_i\right)\right]-r^2\left[\cos\left(\psi-\phi_j\right)-\cos\left(\psi-\phi_i\right)\right]^2}}{r\sin\psi\left(\sin\phi_j-\sin\phi_i\right)+\left(1+r\cos\psi\right)\left(\cos\phi_j-\cos\phi_i\right)}.
\end{equation}
Next, we draw a line $v$ parallel to the $z$ axis passing through $P$: this line acts as the shared boundary of three half-planes, $\pi_{12}$, $\pi_{23}$ and $\pi_{31}$, such that $\pi_{ij}$ contains $v$ and the point $P_{ij}$. It can be shown that the intersection between these half-planes and the sphere identifies three Laguerre cells $C_1,C_2,C_3$, which are wedges with an axis in general different from the axis of the sphere.

	\begin{figure}
		\centering
	\includegraphics[width=0.495\textwidth]{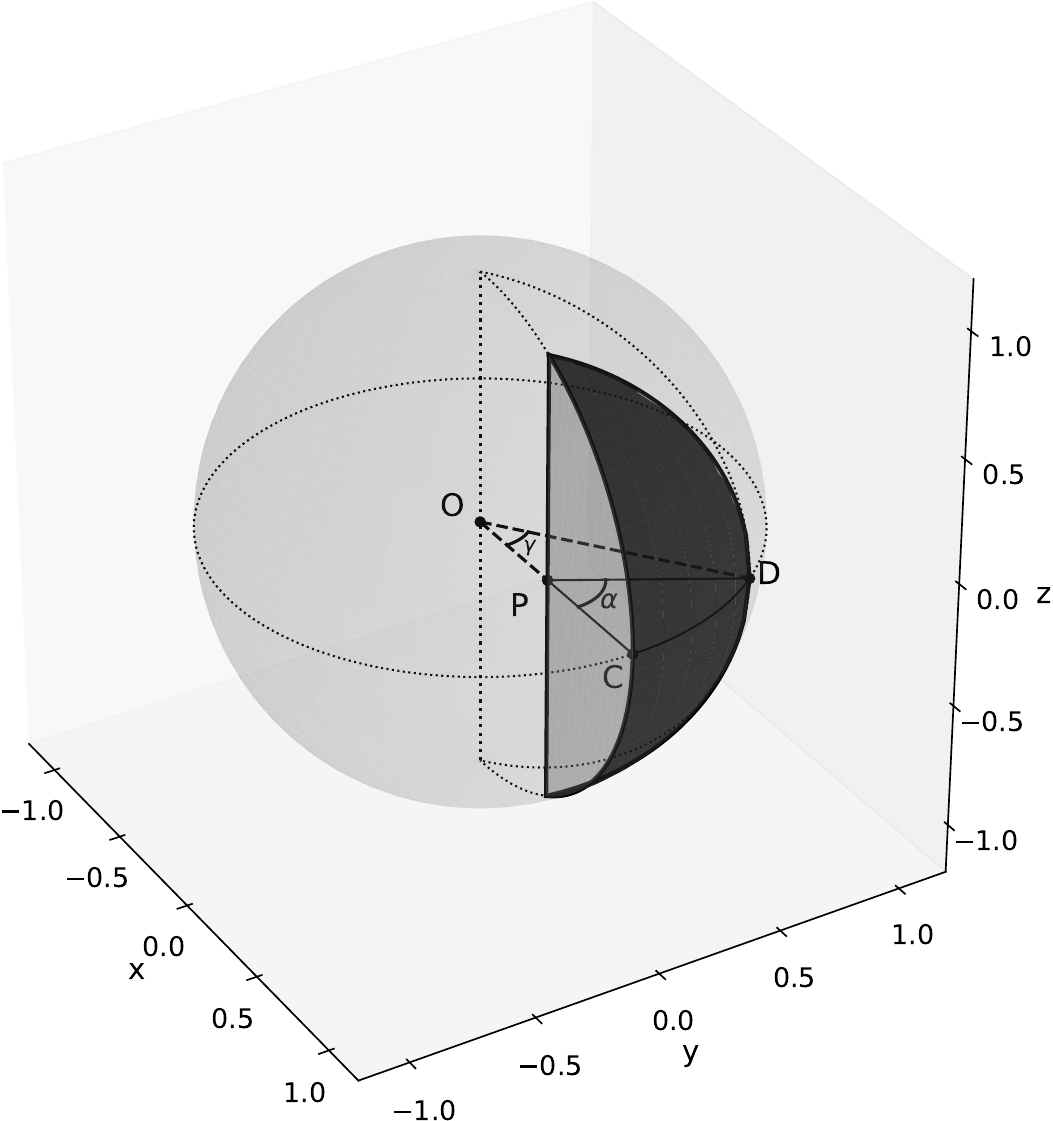}
				\hfill
		\includegraphics[width=0.495\textwidth]{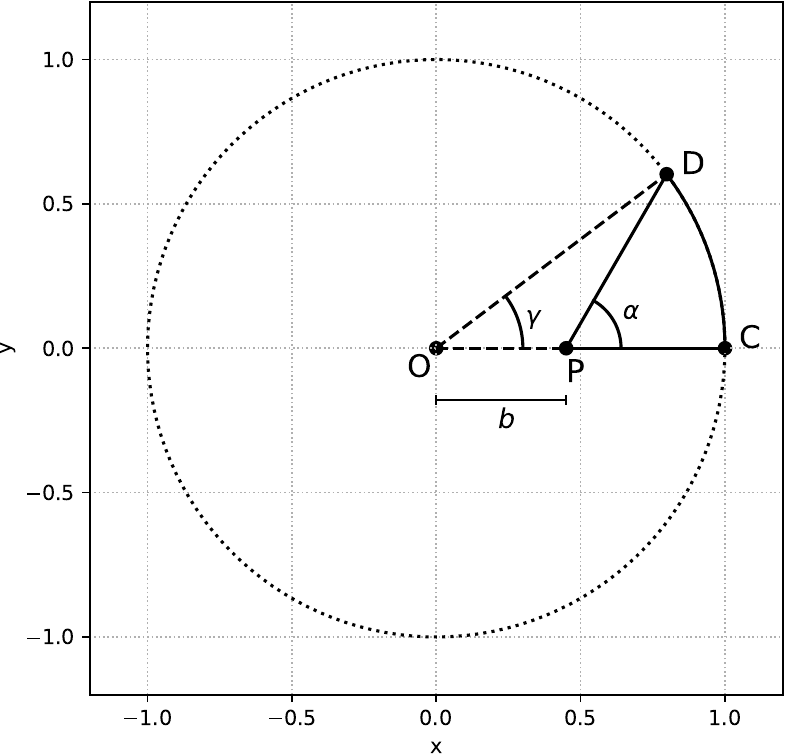}
		\caption{Graphical visualisation of a wedge, whose area is $A\left(b,\alpha\right)$. On the left, the complete three-dimensional wedge on the sphere; on the right, its two-dimensional projection on the $z=0$ plane. $b=0.45$ and $\alpha=\pi/3$ are the values used in this visualisation.}
		\label{wedge}
	\end{figure}
    
	Each choice of the point $P$ leads to a different partition of the sphere into three Laguerre cells. The correct choice is found by imposing that the areas of the wedges, $A_1$, $A_2$ and $A_3$ - which depend on the coordinates of $P$ - are weighted by the energy fractions (eq. (\ref{constraints_semi_D}) and piece-wise transportation plan):
	\begin{equation}
		\label{constraints_3sphere}
		\left\{
		\begin{array}{l}
			A_1\left(r,\psi\right)=2x_1\pi
			\\
			A_2\left(r,\psi\right)=2x_2\pi
		\end{array}
		\right..
	\end{equation}
	
	To proceed, we first need an explicit expression for the area of a wedge. We begin with the simplest configuration (see Fig.~\ref{wedge}): in the plane $z=0$, we consider the point $P(b,0)$, with $0<b<1$, and the point $C(1,0)$; from $P$, we draw a segment forming an angle $\alpha$ with the segment $PC$, with $0<\alpha<\pi$. This segment intersects the circumference at a point $D(\cos\gamma,\sin\gamma)$, where $\gamma$ is the azimuthal angle of $D$ with respect to the centre $O$ of the circle. The relations between $\gamma$ and $\alpha$ are
	\begin{equation}
		\label{gamma_alpha}
		\gamma=\alpha-\arcsin\left(b\sin\alpha\right),\hspace{5ex}\alpha=\arccos\frac{\cos\gamma-b}{\sqrt{1+b^2-2b\cos\gamma}}.
	\end{equation}
	\\
    The area of the wedge is given by the integral
	\begin{equation}
        \label{A_wedge_int}
		A\left(b,\alpha\right)=2\int_{-\pi}^{\pi}d\phi'\,\int_{0}^{1}d\cos\theta'\,\Theta\left(\phi'\right)\Theta\left(\gamma-\phi'\right)\Theta\left(s(\phi')-\cos\theta'\right)=2\int_{0}^{\gamma}d\phi'\,s(\phi').
	\end{equation}
	Here, $s(\phi')$ denotes the cosine of the limiting polar angle $\theta_m(\phi')$: for a fixed azimuth $\phi'\in[0,\alpha]$, the arc of the sphere lying inside the wedge spans the polar interval $\left[\theta_m,\pi-\theta_m\right]$. Thus, $s(\phi')$ is the maximal value of $\cos\theta'$ for which the point $(\theta',\phi')$ still belongs to the wedge:

    \begin{equation}
        s(\phi')=\sqrt{1-\frac{b^2\sin^2\alpha}{\sin^2\left(\alpha-\phi'\right)}}.
    \end{equation}
	Performing the integration in eq.~(\ref{A_wedge_int}) gives
	\begin{equation}
		\label{A_wedge}
		A\left(b,\alpha\right)=2\left[b\sin\alpha\arctan\left(\frac{b\cos\alpha}{\sqrt{1-b^2}}\right)-\arctan\left(\frac{\cot\alpha}{\sqrt{1-b^2}}\right)\right]+\pi\left(1-b\sin\alpha\right).
	\end{equation}

	In order to apply this result to the areas of the Laguerre cells, we must extend its validity beyond the configuration $b>0$ and $0<\alpha<\pi$. We do this by considering signed areas and by imposing the following rules:
	\begin{itemize}
		\item $\alpha$ negative:
		\begin{equation}
			A\left(b,\alpha\right)=-A\left(b,-\alpha\right);
		\end{equation}
		\item $b$ negative (assuming $\alpha$ positive):
		\begin{equation}
			A\left(b,\alpha\right)=2\pi-A\left(-b,\pi-\alpha\right).
		\end{equation}
	\end{itemize}

    Finally, after introducing $\tilde{A}\left(b,\gamma\right)=A\left(b,\alpha(\gamma)\right)$, the expression of the wedge area in terms of $\gamma$ via eq.~(\ref{gamma_alpha}), we can express the areas of the cells in terms of the coordinates of $P$:
    \begin{subequations}
	\begin{align}
		A_1\left(r,\psi\right) & = \tilde{A}\left(r,\xi_{12}-\psi\right)-\tilde{A}\left(r,\xi_{31}-\psi\right),
		\\
		A_2\left(r,\psi\right) & = \tilde{A}\left(r,\xi_{23}-\psi\right)-\tilde{A}\left(r,\xi_{12}-\psi\right),
		\\
		A_3\left(r,\psi\right) & = \tilde{A}\left(r,\xi_{31}-\psi\right)-\tilde{A}\left(r,\xi_{23}-\psi\right),
	\end{align}
    \end{subequations}
    recalling that $\xi_{ij}$ depend on $r,\psi$ according to eq.~(\ref{varphi_ij}).
	It is clear that these three areas sum to zero, meaning that one of them is necessarily negative: this is resolved by adding $4\pi$ to this area, ensuring that all the three wedges have a positive area.

	To conclude this first part of the procedure, we solve the system of simultaneous equations (\ref{constraints_3sphere}) to determine the position of point $P$. This system has no closed-form solution and has to be solved numerically; this is the only non-analytic step in the entire procedure.

    \subsection{Computation of the isotropy}
	After partitioning the sphere into three Laguerre cells, we integrate the distances over each cell according to eq.~(\ref{I_3sph}).
As in the case of the cell areas, the isotropy can be decomposed into wedge-level contributions. Let us consider a particle at angle $\beta$: its contribution to the isotropy, integrated over a wedge defined by $\left(b,\alpha\right)$ lying inside a cell $C_i$ (see Fig.~\ref{wedge}), is given by
	\begin{align}
		\label{I_integral}
		I(b,\alpha,\beta) & = \frac{1}{2\pi}\int_{-\pi}^{\pi}d\phi'\int_{0}^{1}d\cos\theta'\,\Theta\left(\phi'\right)\Theta\left(\gamma-\phi'\right)\Theta\left(s(\phi')-\cos\theta'\right)2\left[1-\sin\theta'\cos\left(\phi'-\beta\right)\right] \nonumber
        \\
        & = \frac{1}{\pi}A\left(b,\alpha\right)-\frac{1}{2\pi}\left\{\sin\beta\arccos{b}-b\sqrt{1-b^2}\sin\alpha\cos\left(\alpha-\beta\right)+\right. \nonumber
		\\
		& \phantom{=} +\left.\sin\left(\alpha-\beta\right)\left(1-b^2\sin^2\alpha\right)\left[\frac{\pi}{2}-\arctan\left(\frac{b\cos\alpha}{\sqrt{1-b^2}}\right)\right]\right\}.
	\end{align}
    As before, we extend the validity of this result by imposing the following rules:
	\begin{itemize}
		
		\item $\alpha$ negative:
		\begin{equation}
			I\left(b,\alpha,\beta\right)=-I\left(b,-\alpha,-\beta\right);
		\end{equation}
		
		\item $b$ negative (assuming $\alpha$ positive):
		\begin{equation}
			I\left(b,\alpha,\beta\right)=1-\frac{1}{2}\sin\beta-I\left(-b,\pi-\alpha,\pi-\beta\right).
		\end{equation}
		
	\end{itemize}
    Next, we introduce $\tilde{I}\left(b,\gamma,\beta\right)=I\left(b,\alpha(\gamma),\beta\right)$, which expresses the wedge isotropy in terms of $\gamma$ through eq.~(\ref{gamma_alpha}). Using this definition, the contributions $I_i$ of the three particles to the isotropy read:
    \begin{subequations}
    \begin{align}
		I_1\left(x_1,x_2\right) & = \tilde{I}\left(r,\xi_{12}-\psi,\phi_1-\psi\right)-\tilde{I}\left(r,\xi_{31}-\psi,\phi_1-\psi\right),
		\\
		I_2\left(x_1,x_2\right) & = \tilde{I}\left(r,\xi_{23}-\psi,\phi_2-\psi\right)-\tilde{I}\left(r,\xi_{12}-\psi,\phi_2-\psi\right),
		\\
		I_3\left(x_1,x_2\right) & = \tilde{I}\left(r,\xi_{31}-\psi,\phi_3-\psi\right)-\tilde{I}\left(r,\xi_{23}-\psi,\phi_3-\psi\right).
	\end{align}
    \end{subequations}
	Since one of the three cell areas is negative, and the isotropy of a one-particle event is 2, we add 2 to the isotropy contribution associated with the cell of negative area.
    
	The total isotropy is then given by the sum of these three contributions:
	\begin{equation}
		\label{I_x1x2_3sph}
\mathcal{I}_{3}\left(x_1,x_2\right)=I_1\left(x_1,x_2\right)+I_2\left(x_1,x_2\right)+I_3\left(x_1,x_2\right).
	\end{equation}
    The final result depends only on the kinematics of the event, since determining the Laguerre cells fixes the dependence of $r$, $\psi$ and $\xi_{ij}$ on the variables $x_1,x_2$.

\subsection{Fully analytic isotropy}
    \label{fully-analytical}

    While the procedure described above is only semi-analytic, fully analytic results for the isotropy value can be obtained for specific configurations. One such case is the fully symmetric three-particle configurations, with the particles at the vertices of an equilateral triangle and with equal energies. This configuration is also the one that gives the largest value of $\iota$. 
     In this case, we have $x_1=x_2=x_3=2/3$ and $\phi_1=0$, $\phi_2=2\pi/3$, $\phi_3=-2\pi/3$. By symmetry, the Laguerre cells -- which in this configuration reduce to Voronoi cells -- are three spherical wedges bounded by $\xi_{12}=\pi/3$, $\xi_{23}=\pi$, $\xi_{31}=-\pi/3$, with their axes aligned with the axis of the sphere. Each particle contributes equally to the isotropy, and the final result is
	\begin{equation}
		\iota_3^\text{max}=1-\mathcal{I}_{3}=1-\frac{3}{2\pi}\int_{-\frac{\pi}{3}}^{\frac{\pi}{3}}d\phi'\int_{-1}^{1}d\cos\theta'\left(1-\sin\theta'\cos\phi'\right)=\frac{3\sqrt{3}}{4}-1 \simeq 0.299038.
	\end{equation}

    Similar symmetry considerations allow one to calculate fully analytically the result for the configuration consisting of four particles of equal energy at the vertices of a regular tetrahedron, the maximally isotropic distribution of four particles. The four Voronoi cells are now identical spherical triangles, each one of area $\pi$. The result is
    \begin{eqnarray}
        \iota_4^\text{max} &=& 1-\mathcal{I}_{4} = 1- \left[2-\frac{12}{\pi}\int_0^{\pi/3}d\phi'\int_0^{\arctan(\sqrt{2}/\cos(\phi'))} d\theta' \cos\theta'\sin\theta' \right] \nonumber \\
       &=&\frac{2\sqrt{6} \arctan(\sqrt{2})}{\pi} -1 \simeq 0.489715.
    \end{eqnarray}
Both the $n=3$ and the $n=4$ values are compatible with the respective upper endpoint for $\iota$ observed in Fig.~\ref{fig:iota_distribution}.

\section{Details of the resummation}
	\label{resum_details}    
In this appendix, we collect the details of the derivation of the resummed result, eq.~(\ref{eq:cumulative-nll}). As customary for $e^+e^-$ collisions, we choose the centre-of-mass energy $\sqrt{Q^2}$ as reference hard scale.
A general framework to achieve NLL resummation for global event shapes was established in~\cite{Banfi:2004yd}. 
The formalism requires two ingredients: the radiator $R$, which accounts for the exponentiation of the one-emission contribution with running coupling, and a correction factor that takes into account the effect of multiple emissions.

\subsection{The radiator}\label{app:radiator}
Following~\cite{Banfi:2004yd}, the radiator is defined as 
\begin{equation}\label{eq:rad-def}
R(\iota)\equiv\sum_{l=1}^{2}\int\left[dk\right]\left|\mathcal{M}^2_{(l)}(k)\right|\Theta\left(V(k)-\iota\right),
	\end{equation}
where the sum is over the two hard legs, i.e. the quark and the antiquark. The matrix element and the phase-space are
	\begin{equation}
		\left|\mathcal{M}^2_{(l)}(k)\right|=\frac{\as \cf}{2\pi}\frac{zp_{qg}(z)}{k_t^2},\hspace{10ex}\left[dk\right]=\frac{dz}{z}\frac{d\phi}{2\pi}dk_t^2,
	\end{equation}
where $z$ and $k_t$ are the emitted gluon energy fraction and transverse momentum with respect to each of the emitting legs and $p_{gq}$ is the time-like splitting function (see~\cite{Banfi:2004yd} for details).

The integrals in eq.~(\ref{eq:rad-def}) can be calculated once the scaling of the observable in the soft and collinear limit is known. Such scaling is written in terms of the emission transverse momentum and rapidity with respect to each hard parton $l$:
\begin{equation}\label{eq:caesar-param}
    V(k)= d_l \, g_l(\phi) \left(\frac{k_t}{Q} \right)^a e^{- b_l \eta^{(l)}}.
\end{equation}
The behaviour of the observable $\iota$ in the soft and collinear limit was determined in sec.~\ref{sec:res}. We have
\begin{equation}
    a=1, \quad b_l=0, \quad d_l=2, \quad g_l(\phi)=1, \quad l=1,2.
\end{equation}
The resulting radiator can be separated into leading and next-to-leading contributions
\begin{equation}
    R(\iota)= R_\text{LL}(\iota)+R_\text{NLL}(\iota),
\end{equation}
where 
\begin{subequations}
\begin{align}
 R_\text{LL}(\iota)&=   -\frac{\cf}{\as\pi\beta_0^2}\left[2\lambda+\log\left(1-2\lambda\right)\right],\\
 R_\text{NLL}(\iota)&= \frac{\cf}{\pi\beta_0}\left\{-2B_q\log\left(1-2\lambda\right)+2 \log 2 \frac{2 \lambda}{1-2\lambda}+\frac{K}{2\pi\beta_0}\left[\log\left(1-2\lambda\right)+\frac{2\lambda}{1-2\lambda}\right]\right. \nonumber
		\\
		& \phantom{=}\left.-\frac{\beta_1}{\beta_0^2}\left[\frac{1}{2}\log^2\left(1-2\lambda\right)+\frac{\log\left(1-2\lambda\right)+2\lambda}{1-2\lambda}\right]
        + \log \frac{\mu_R^2}{Q^2}\left[\log(1-2 \lambda)+\frac{2 \lambda}{1-2 \lambda} \right]\right\},
\end{align}
\end{subequations}
where $\lambda= -\as \beta_0 \log \iota$, with $\as=\as(\mur^2)$. We have also introduced the quark and gluon colour factors: $\cf=4/3$ and $\ca=3$, and the quark hard-collinear constant $B_q=-\frac{3}{4}$. The above results have been obtained with the running of the strong coupling at two loops in the so-called CMW scheme~\cite{Catani:1990rr}:
\begin{equation}
		\as^{\text{CMW}}(k_t^2)=\as(k_t^2) \left[ 1+ \frac{\as(k_t^2)}{2\pi}K
        \right],
	\end{equation}
where $\as(k_t^2)$ is the standard $\msb$ coupling at two-loop accuracy:
\begin{equation}
\as\left(k_t^2\right)=\frac{\as(\mur^2)}{1+\tilde{\lambda}}\left(1-\as(\mur^2)\frac{\beta_1}{\beta_0}\frac{\log\left(1+\tilde{\lambda}\right)}{1+\tilde{\lambda}}\right),\hspace{10ex}\tilde{\lambda}=\as(\mur^2)\beta_0\log\frac{k_t^2}{\mur^2},
\end{equation}
with
\begin{equation}\label{eq:beta-function}
		\beta_0=\frac{11\ca-2\nf}{12\pi},\hspace{10ex}\beta_1=\frac{17\ca^2-5\ca\nf-3\cf \nf}{24\pi^2},
	\end{equation}
and
\begin{equation}\label{eq:cmw-K}
K=\ca\left(\frac{67}{18}-\frac{\pi^2}{6}\right)-\frac{5}{9}\nf.
\end{equation}

\subsection{The resummation constant $C_1$}\label{app:C1}
We now describe the analytic calculation of the resummation constant $C_1$ that appears in eq.~(\ref{eq:cumulative-nll}).
In particular, we follow the approach taken, for instance, in~\cite{Ghira:2024nkk} and we compute the differential distribution for $\iota\ge0$, i.e.\ including the endpoint contribution at $\iota=0$, from which the constant $C_1$ can be read off. To this purpose, we must consider both real-emission and virtual corrections in dimensional regularisation. 
We start quoting the result for the latter. Working in $d=4-2\epsilon$ dimensions, the $\msb$ renormalised virtual contribution reads: 
\begin{align}\label{eq:virtual}
    \frac{1}{\sigma^{(d)}_0}\frac{d \sigma^{(V)}}{d \iota}= \frac{\as \cf S_\epsilon}{2\pi} \left[-\frac{2}{\epsilon^2} -\frac{3}{\epsilon} -8 + \pi^2 \right]\delta(\iota),
\end{align} 
where, for convenience, we have defined $S_\epsilon= \left(\frac{4\pi \mu_R^2}{Q^2}\right)^\epsilon\frac{1}{\Gamma(1-\epsilon)}$ and denoted with $\sigma_0^{(d)}$ the Born level massless cross section in $d$ dimensions.

\begin{figure}
    \centering
\includegraphics[width=0.49\textwidth,page=1]{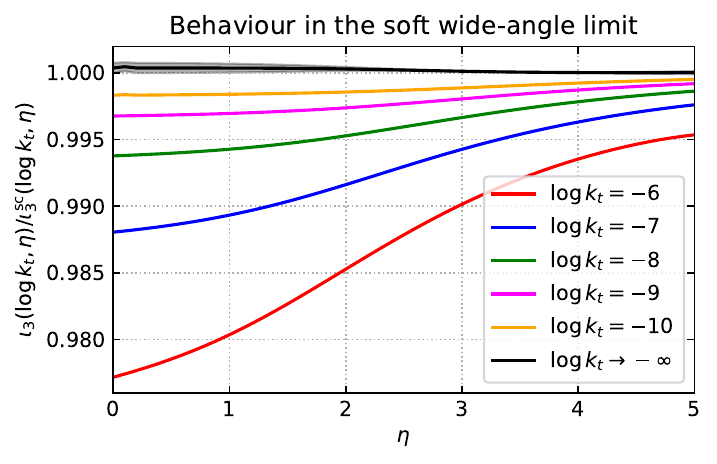}
\includegraphics[width=0.49\textwidth,page=2]{figures/iota-ir-behaviour.pdf}
	\caption{Ratio of the (three-particle) isotropy $\iota$ computed with full kinematics over its soft and collinear approximation $\iota_3^\text{sc}$, as a function of either the emission rapidity $\eta$ (left plot), or of the momentum fraction $z$ (right plot), for different values of the transverse momentum $\kt$.}
		\label{fig:iota-behaviour}
\end{figure}

Next, we have to consider the contribution from the real emission. Because we want to capture the delta-contribution, we must lift the soft and collinear assumption of the NLL calculation. In particular, we have to consider the behaviour of event isotropy in the presence of a hard collinear or soft-wide angle emission. Because we cannot rely on the soft-and-collinear behaviour derived analytically in eq.~(\ref{soft-momenta}), we employ a numerical study. 
In Fig.~\ref{fig:iota-behaviour}, we plot the ratio of the three-particle observable $\iota_3$, computed with full kinematics, over its soft and collinear approximation, see eq.~(\ref{eq:caesar-param}), as a function of the emission rapidity, and for different values of $\kt$, on the left, or as a function of the momentum fraction $z= \frac{\kt}{Q}e^{-\eta}$.
The plot on the left allows us to study the effect of a soft emission at wide angle, i.e.\ small $\eta$, while the one on the right sheds light on the full $z$ dependence, including the hard-collinear region.
The plots also contain an extrapolation to vanishing $\kt$.
In all cases, as $\log \kt$ becomes large and negative, i.e. $\kt$ grows smaller, both ratios approach unity. 
Thus, we conclude that the behaviour $\iota_3^\text{sc}= \frac{2\kt}{Q}$ holds in both the soft-wide angle and hard-collinear regions, and we do not have to consider the two contributions separately.

The calculation described in~\cite{Ghira:2024nkk} cannot be directly applied to the case of an observable that scales like a transverse momentum, such as $\iota$. For this reason, we perform the calculation for the more generic angularity
\begin{equation}\label{eq:lambda_def}
\lambda_\alpha= \frac{\kappa^\alpha}{[z(1-z)]^{\alpha-1}},
\end{equation}
with $\kappa=\sqrt{\frac{k_t^2}{Q^2}}$. Note that the limit $\alpha\to1$ is smooth and can be taken at the end of the calculation, leading to $\iota=2\lambda_1$.
The real emission contribution can be written as
\begin{align}\label{eq:real-start}
    \frac{1}{\sigma^{(d)}_0}\frac{d \sigma^{(R)}}{d \lambda_\alpha}&= S_\epsilon \int_0^1 d z \int_0^{z^2 (1-z)^2} \frac{d \kappa^2}{\kappa^{2+ 2\epsilon}} P_{gq}(z,\epsilon) \,\delta \left(\kappa^{\alpha} z^{1-\alpha}(1-z)^{1-\alpha} -  \lambda_\alpha \right) \nonumber\\
    &=\frac{\as \cf S_\epsilon }{\pi}  \int_0^1 d z \int_0^{z^2 (1-z)^2} \frac{2 d \kappa}{\kappa^{1+ 2\epsilon}} \left(\frac{2}{z}-2+(1-\epsilon) z \right) \delta \left(\kappa^{\alpha} z^{1-\alpha}(1-z)^{1-\alpha} -  \lambda_\alpha \right) \nonumber\\
    &=\frac{\as \cf S_\epsilon}{\pi} \frac{2}{\alpha \, \lambda_\alpha^{1+\frac{2 \epsilon}{\alpha}}}  \int_{z_{\text{min}}}^{z_{\text{max}}} d z z^{2 \epsilon \frac{1-\alpha}{\alpha} }(1-z)^{2 \epsilon \frac{1-\alpha}{\alpha} }  \left(\frac{2}{z}-2+(1-\epsilon) z \right), 
\end{align}   
where we have considered the two (identical) contributions arising from the gluon being emitted collinear to the quark or to the antiquark. 
Up to power corrections in $\lambda_\alpha$, that we neglect, we can take $z_{\text{min}}=\lambda_\alpha$ and $z_{\text{max}}=1$, arriving at
\begin{align}\label{eq:real-cntd}
    \frac{1}{\sigma^{(d)}_0}\frac{d \sigma^{(R)}}{d \lambda_\alpha}&=\frac{\as \cf S_\epsilon}{\pi} \frac{1}{\alpha} \frac{2}{ \lambda_\alpha^{1+\frac{2 \epsilon}{\alpha}}} \nonumber
    \\
    & \quad \times\int_{\lambda_\alpha}^{1} d z z^{2 \epsilon \frac{1-\alpha}{\alpha} } \left[1+2 \epsilon \frac{1-\alpha}{\alpha} \log (1-z)+ \mathcal{O}(\epsilon^2)\right]\left(\frac{2}{z}-2+(1-\epsilon) z \right).  
\end{align}   
The integral over the momentum fraction $z$ can be done using the well-known identity
\begin{align}
 \frac{1}{\lambda^{1+a \epsilon}}= -\frac{1}{a \epsilon}\delta(\lambda) + \left(\frac{1}{\lambda}\right)_+- a \epsilon \left(\frac{\log \lambda}{\lambda}\right)_+ +\mathcal{O}(\epsilon^2).
\end{align}
The first contribution is
\begin{align}
\frac{1}{\alpha} \frac{2}{ \lambda_\alpha^{1+\frac{2 \epsilon}{\alpha}}}  \int_{\lambda_\alpha}^{1} d z z^{2 \epsilon \frac{1-\alpha}{\alpha} }\frac{2}{z}&=\frac{1}{1-\alpha}\frac{2
}{\epsilon}\left( \frac{1}{ \lambda_\alpha^{1+\frac{2 \epsilon}{\alpha}}} - \frac{1}{ \lambda_\alpha^{1+2 \epsilon}}  \right)\nonumber \\
&=\frac{1}{1-\alpha}\frac{2}{\epsilon} \left (\frac{1-\alpha}{2 \epsilon} \delta(\lambda_\alpha)- 2 \epsilon \frac{1-\alpha}{\alpha} \left(\frac{\log \lambda_\alpha}{\lambda_\alpha} \right)_+\right)+ \mathcal{O}(\epsilon)\nonumber\\
&=\frac{1}{\epsilon^2} \delta(\lambda_\alpha)-\frac{4}{\alpha} \left(\frac{\log \lambda_\alpha}{\lambda_\alpha} \right)_+ + \mathcal{O}(\epsilon).
\end{align}   
For the finite part of the splitting function, we can safely set the lower limit of integration to zero. We have
\begin{align}
\frac{1}{\alpha} \frac{2}{ \lambda_\alpha^{1+\frac{2 \epsilon}{\alpha}}}  \int_{0}^{1} d z z^{2 \epsilon \frac{1-\alpha}{\alpha} }(-2 +(1-\epsilon) z)=
\frac{3}{2\epsilon} \delta(\lambda_\alpha) -\frac{3}{\alpha} \left(\frac{1}{\lambda_\alpha} \right)_+ - \frac{7-8 \alpha}{2 \alpha}\delta(\lambda_\alpha) +\mathcal{O}(\epsilon).
\end{align}   
In summary, the real-emission contribution reads
\begin{align}
    \frac{1}{\sigma^{(d)}_0}\frac{d \sigma^{(R)}}{d \lambda_\alpha}&=
    \frac{\as \cf S_\epsilon}{2\pi} \Bigg[ 
    \left(\frac{2}{\epsilon^2} + \frac{3}{\epsilon}  - \frac{7-8 \alpha}{\alpha} \right)\delta(\lambda_\alpha) -\frac{8}{\alpha} \left(\frac{\log \lambda_\alpha}{\lambda_\alpha} \right)_+   -\frac{6}{\alpha} \left(\frac{1}{\lambda_\alpha} \right)_+\Bigg].
\end{align}

Let us sum together the virtual and the real contributions. This is enough to cancel the poles and reproduce the logs. The poles cancel for any value of $\alpha$, as they should. We can then set $\epsilon = 0$ and, for $\alpha=1$, we obtain
\begin{align}
    \frac{1}{\sigma^{(d)}_0}\frac{d \sigma}{d \lambda_1}&= \frac{\as \cf}{2\pi} \ \Bigg[ - 8 \left(\frac{\log \lambda_1}{\lambda_1} \right)_+ -6 \left(\frac{1}{\lambda_1} \right)_+ + \left( \pi^2-7\right) \delta(\lambda_1)
    \Bigg].
\end{align}
This leads to the cumulative distribution
\begin{align}
\Sigma (\iota)&= 1+ \frac{\as \cf }{2\pi} \ \left[ - 4 \log^2 \frac{\iota}{2} -6 \log \frac{\iota}{2} + \pi^2-7 \right] \nonumber\\
&=  \frac{\as \cf }{2\pi} \left[ - 4 \log^2 \iota+ (8 \log 2-6) \log \iota + \pi^2-7-4 \log^2 2+ 6 \log2  \right].
\end{align}
From this expression, we can read off
\begin{equation}\label{eq:C1-result}
    C_1=  \pi^2-7-4 \log^2 2+ 6 \log2.
\end{equation}

\subsection{Multiple-emission behaviour}\label{app:ME}
In this section we study the additivity of $\iota$ both numerically and analytically. Starting from two hard partons with four-momenta $p_{1,2}$, we consider two soft-and-collinear emissions $p_3$ and $p_4$ with transverse momentum and rapidity $k_{t{3,4}}$ and $\eta_{3,4}$, respectively. Let $\phi_{34}$ be the azimuthal angle between the two emitted partons. We assume that the two partons are emitted in the same hemisphere, meaning that $\eta_3$ and $\eta_4$ have the same sign.
We call $\iota^\text{sc}_4(p_3,p_4)$ the isotropy of this soft-collinear four-particle configuration, while we call $\iota^\text{sc}_3(p_3)$ and $\iota^\text{sc}_3(p_4)$ the isotropy of the configurations with one soft-collinear emission only, particle 3 and 4 respectively.
The additivity hypothesis is
\begin{equation}
    \iota^\text{sc}_4(p_3,p_4)=\iota^\text{sc}_3(p_3) + \iota^\text{sc}_3(p_4).
\end{equation}
To prove this analytically, we consider the kinematic variables $z_{3,4}$ and $\theta_{3,4}$, respectively energy fractions and polar angles of the soft partons. In the limit of small $z$ and $\theta$ \footnote{From now on, $z$ is used as a shorthand for either $z_3$ or $z_4$, and likewise for $\theta$. For instance, $\mathcal{O}\left(z\theta^2\right)$ may stand for $\mathcal{O}\left(z_3\theta_3^2\right)$, $\mathcal{O}\left(z_4\theta_4^2\right)$, or $\mathcal{O}\left(z_3\theta_3^2\right)+\mathcal{O}\left(z_4\theta_4^2\right)$. Note that the definition of $z$ used here coincides with the definition introduced in eq. (\ref{z_theta_def}) only for small $z$.}, exploiting the azimuthal symmetry the four-momenta can be written as
\begin{subequations}
    \begin{align}
    p_1 & = \frac{Q}{2}\left(1-z_3-z_4,-z_3\theta_3-z_4\theta_4\cos\phi_{34},-z_4\theta_4\sin\phi_{34},1-z_3-z_4\right),
    \\
    p_2 & = \frac{Q}{2}\left(1,0,0,-1\right),
    \\
    p_3 & = \frac{Q}{2}\left(z_3,z_3\theta_3,0,z_3\right),
    \\
    p_4 & = \frac{Q}{2}\left(z_4,z_4\theta_4\cos\phi_{34},z_4\theta_4\sin\phi_{34},z_4\right).
    \end{align}
\end{subequations}
The distances of the points of the sphere from the particles read
\begin{subequations}
    \begin{align}
    d_1\left(\theta',\phi'\right) & = 2\left[1+\left(z_3\theta_3+z_4\theta_4\cos\phi_{34}\right)\sin\theta'\cos\phi'+z_4\theta_4\sin\phi_{34}\sin\theta'\sin\phi'\right. \nonumber
    \\
    & \phantom{=} -\left.\cos\theta'\right]+\mathcal{O}\left(z^2\theta\right)+\mathcal{O}\left(z\theta^2\right),
    \\
    d_2\left(\theta',\phi'\right) & = 2\left[1+\cos\theta'\right],
    \\
    d_3\left(\theta',\phi'\right) & = 2\left[1-\theta_3\sin\theta'\cos\phi'-\cos\theta'\right]+\mathcal{O}\left(z^2\theta\right)+\mathcal{O}\left(z\theta^2\right),
    \\
    d_4\left(\theta',\phi'\right) & = 2\left[1-\theta_4\cos\phi_{34}\sin\theta'\cos\phi'-\theta_4\sin\phi_{34}\sin\theta'\sin\phi'\right. \nonumber
    \\
    & \phantom{=} -\left.\cos\theta'\right]+\mathcal{O}\left(z^2\theta\right)+\mathcal{O}\left(z\theta^2\right).
    \end{align}
\end{subequations}

As done for one emission, we consider the Born level configuration. With just the two hard partons $\bar 1$ and $\bar 2$, we have $\iota=0$ and the two Laguerre cells are the Northern and Southern hemispheres: $C_{\bar 1}= \Sigma_N$ and $C_{\bar 2}= \Sigma_S$, with $\Sigma_N=\left\{(\theta',\phi'):\theta'\in[0,\pi/2)\right\}$ and $\Sigma_S=\left\{(\theta',\phi'):\theta'\in[\pi/2,\pi)\right\}$. In the Born configuration, $d_{\bar{1}}\left(\theta',\phi'\right)=2\left[1-\cos\theta'\right]$, while $d_{\bar{2}}\left(\Omega'\right)=d_2\left(\Omega'\right)$.
Starting from eq. (\ref{I_nsph}), adding and subtracting the Born level isotropy, and with  algebraic manipulations very similar to those used to obtain eq.~(\ref{is3-sc-cntd}), we obtain
\begin{align}
    \label{i_sc_double_34}
    1-\iota_{4}^\text{sc}(p_3,p_4) & = 1+\frac{1}{4\pi}\int_{C_2\setminus\Sigma_S}d\Omega'\,\left(d_2^2\left(\Omega'\right)-d_1^2\left(\Omega'\right)\right)+\frac{1}{4\pi}\int_{\Sigma_N}d\Omega'\,\left(d_1^2\left(\Omega'\right)-d_{\bar{1}}^2\left(\Omega'\right)\right) \nonumber
    \\
    & \phantom{=} +\frac{1}{4\pi}\int_{C_3}d\Omega'\,\left(d_3^2\left(\Omega'\right)-d_1^2\left(\Omega'\right)\right)+\frac{1}{4\pi}\int_{C_4}d\Omega'\,\left(d_4^2\left(\Omega'\right)-d_1^2\left(\Omega'\right)\right).
\end{align}
The first two integrals in eq. (\ref{i_sc_double_34}) give a sub-leading contribution:
\begin{subequations}
\begin{align}
    \int_{\Sigma_N}d\Omega'\,\left(d_1^2\left(\Omega'\right)-d_{\bar{1}}^2\left(\Omega'\right)\right) & = \mathcal{O}\left(z^2\theta\right)+\mathcal{O}\left(z\theta^2\right),
    \\
    \label{d2-d1_int}
    \int_{C_2\setminus\Sigma_S}d\Omega'\,\left(d_2^2\left(\Omega'\right)-d_1^2\left(\Omega'\right)\right) & = \mathcal{O}\left(z^2\theta\right).
\end{align}
\end{subequations}
Eq. (\ref{d2-d1_int}) is obtained assuming that $C_2\setminus\Sigma_S$ is a neighbourhood of $\theta'=\pi/2$. The last two integrals are estimated considering that $C_3$ is a neighbourhood of $\left(\theta',\phi'\right)=\left(\pi/2,0\right)$, with area $A_3=2\pi{}z_3$, and $C_4$ is a neighbourhood of $\left(\theta',\phi'\right)=\left(\pi/2,\phi_{34}\right)$, with area $A_4=2\pi{}z_4$:
\begin{subequations}
\begin{align}
\int_{C_3}d\Omega'\,\left(d_3^2\left(\Omega'\right)-d_1^2\left(\Omega'\right)\right) & = -4\pi{}z_3\theta_3+\mathcal{O}\left(z^2\theta\right)+\mathcal{O}\left(z^2\theta^2\right).
\\
\int_{C_4}d\Omega'\,\left(d_4^2\left(\Omega'\right)-d_1^2\left(\Omega'\right)\right) & = -4\pi{}z_4\theta_4+\mathcal{O}\left(z^2\theta\right)+\mathcal{O}\left(z^2\theta^2\right).
\end{align}
\end{subequations}
Finally, we obtain
\begin{equation}
    \iota_4^\text{sc}(p_3,p_4)=z_3\theta_3+z_4\theta_4 = \iota_3^\text{sc}(p_3) +\iota_3^\text{sc}(p_4)  ,
\end{equation}
which proves the additivity.

\begin{figure}
    \centering
\includegraphics[width=\textwidth]{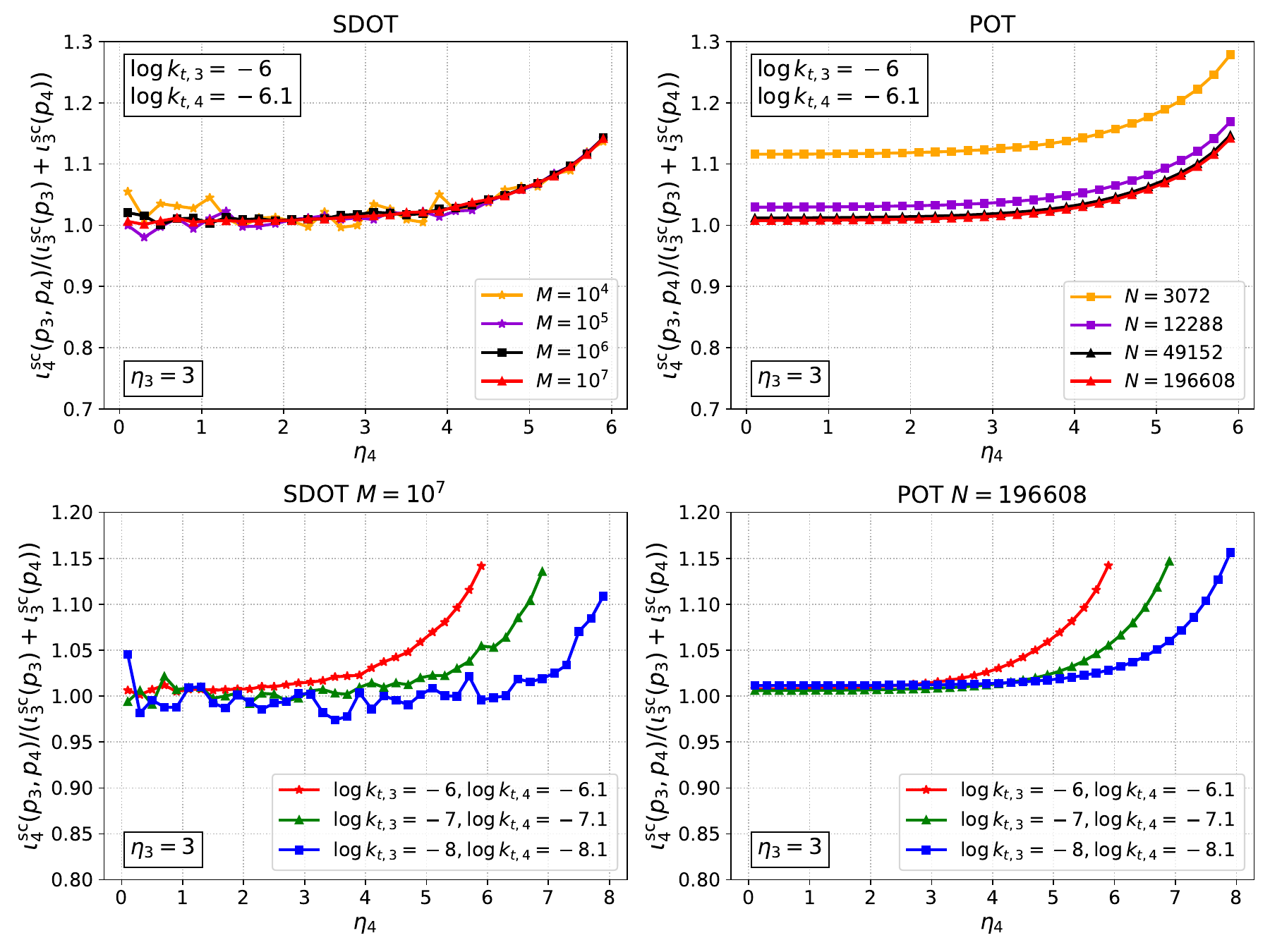}
	\caption{Ratio $\frac{\iota^\text{sc}_4(p_3,p_4)}{\iota^\text{sc}_3(p_3)+\iota^\text{sc}_3(p_4)}$ as a function of the rapidity $\eta_4$, computed with full kinematics using both SDOT and POT.} 
		\label{fig:test_additivity}
\end{figure}
A numerical test of this additivity is shown in Fig.~\ref{fig:test_additivity}. The plots on the left are obtained with the code SDOT, the ones on the right with POT.
We consider the emission of two soft gluons, 3 and 4 with similar transverse momenta, and we fix, for definiteness, the rapidity $\eta_3=3$. We show the ratio  $\frac{\iota^\text{sc}_4(p_3,p_4)}{\iota^\text{sc}_3(p_3)+\iota^\text{sc}_3(p_4)}$ as a function of the rapidity $\eta_4$.

In the top plots we also fix the transverse momenta of the two emissions to be $\log \kt \simeq -6$, and we show curves corresponding to different values of the parameter that controls the accuracy of the solution to the OT problem, $M$ and $N$, respectively. 
While these plots do confirm the property of additivity, they also show that achieving a numerical verification requires large values of $M$ or $N$ and, consequently, expensive calculations. 

In the bottom plots we fix instead $M=10^7$ and $N=196608$, while studying different values of the transverse momenta.
Note that these values of $M$ and $N$ would make the calculation of fixed-order distributions prohibitively expensive.

The plots also provide a qualitative explanation for the inclusive numerical checks of the $\ord{\as^2}$ single-logarithmic contribution in Fig.~\ref{fig:smalliota}. In the region $\log \iota \simeq \log \kt \simeq -6$, we find that SDOT with $M=10^5$ or POT with $N=3072$ captures additivity only to about $10\%$ accuracy, thereby altering the logarithmic structure.

\addcontentsline{toc}{section}{References}

\bibliographystyle{jhep}
\bibliography{biblio}
\end{document}